\documentclass[final,floatfix,twocolumn,pre,aps,floats,amsfonts,amssymb,showpacs]{revtex4}
\usepackage{graphicx}
\usepackage{bm}
\usepackage{longtable}

\def\EQ{\begin{equation}}
\def\EN{\end{equation}}
\def\EQA{\begin{eqnarray}}
\def\ENA{\end{eqnarray}}

\def\bdes{\begin{description}}
\def\edes{\end{description}}

\begin{document}

\title{Stability and turbulent transport in rotating shear flows: \\ prescription from analysis of cylindrical and plane Couette flows data}
\author{B. Dubrulle, O. Dauchot, F. Daviaud}
\affiliation{CNRS URA 2464 GIT/SPEC/DRECAM/DSM, CEA Saclay, F-91191 Gif-sur-Yvette, France}
\author{P-Y. Longaretti}
\affiliation{LAOG UMR 5571 CNRS Universit\'e J. Fourier, F-38041 Grenoble, France}
\author{D. Richard, J-P. Zahn}
\affiliation{LUTh CNRS UMR 8102, Observatoire de Paris-Meudon, Place Jules Janssen, F-92195 Meudon, France}

\begin{abstract}
This paper provides a prescription for the turbulent viscosity in rotating shear flows for use e.g. in geophysical and astrophysical contexts. This prescription is the result of the detailed analysis of the experimental data obtained in several studies of the transition to turbulence and turbulent transport in Taylor-Couette flow. We first introduce a new set of control parameters, based on dynamical rather than geometrical considerations, so that the analysis applies more naturally to rotating shear flows in general and not only to Taylor-Couette flow. We then investigate the transition thresholds in the supercritical and the subcritical regime in order to extract their general dependencies on the control parameters. The inspection of the mean profiles provides us with some general hints on the mean to laminar shear ratio. Then the examination of the torque data allows us to propose a decomposition of the torque dependence on the control parameters in two terms, one completely given by measurements in the case where the outer cylinder is at rest, the other one being a universal function provided here from experimental fits. As a result, we obtain a general expression for the turbulent viscosity and compare it to existing prescription in the literature. Finally, throughout all the paper we discuss the influence of additional effects such as stratification or magnetic fields.
\vspace{1cm}
\end{abstract}

\date{Physics of Fluids, vol 17 , 095103 (2005)}
\maketitle

\section{Introduction}

\label{intro}

One of the basic principle of fluid mechanics is the so-called "Reynolds similarity principle": no matter their composition, size, nature, different flow obeying the same equations with the same control parameters will follow the same dynamics. This principle has been used a lot in engineering to built e.g. prototypes of bridges to be tested in wind tunnels before construction. To obtain easy-to-use prototypes with realistic control parameters, one then decreases the size but increases the velocity of the in-flowing wind so as to keep constant the Reynolds number, controlling the dynamics of the flow. This principle could also be of great interest for certain astrophysical flows, whose dynamics could well be approached by simple laboratory flows. A good example is circum-stellar disk. In \cite{HDH03}, it has been shown that under simple, but founded approximations, their equation of motions were similar to the equation of motion of an incompressible rotating shear flow, with penetrable boundary conditions and cylindrical geometry. This kind of flow can be achieved in the Couette-Taylor flow, a fluid layer sheared between two coaxial cylinders rotating at different speed, while penetrable boundary conditions can be obtained using porous material. On more general grounds, the Taylor-Couette device is also an excellent prototype to study transport properties of most astrophysical or geophysical rotating shear flows: depending on the rotation speed of each cylinder, one can obtain various flow regimes with increasing or decreasing angular velocity and/or angular momentum. 

The Taylor-Couette flow is a classical example of simple system with complex and rich stability properties, as well as prototype of anisotropic, inhomogeneous turbulence. It has therefore motivated a great amount of laboratory experiments, and is even the topic of a major international conference. Tagg (see http://carbon.cudenver.edu/\~rtagg) has conducted a bibliography on Taylor-Couette flow, which gives a good idea of the prototype status of this flow.

Here, we make use of the many results obtained so far for the Taylor-Couette experiment, regarding transition to turbulence, or turbulence properties to propose a practical prescription for the turbulent viscosity as a function of the radial position and the control parameters. It reads 
\begin{equation}
\nu_t=\frac{1}{2\pi}{R_{\cal C}}^4\frac{G_i(Re,\eta )}{Re^2}h(R_\Omega,\eta )\frac{S_{lam}}{\bar S}{\tilde S}{\tilde r}^2,
\end{equation}
where $Re, R_\Omega, R_{\cal C}$ are the control parameters (Section \ref{contro}), $G_i(Re, \eta)$ is the torque measured when only the inner cylinder is rotating (Section \ref{torque-out}), $h(R_\Omega,\eta )$ is a universal function provided in Section \ref{torque-out}, and $S_{lam}/{\bar S}$ is the ratio of the laminar to the mean shear, which encodes all the radial dependence as illustrated in Section \ref{torque-out}.  ${\tilde S}$ and ${\tilde r}$ are typical shear and radius of the considered flow. Most of the results we use here have been published elsewhere, except recent experimental results obtained by Richard~\cite{Rich01}. Our work therefore completes and generalizes the approach pioneered by Zeldovich~\cite{Zel81}, with subsequent contributions by ~\cite{Bulle93,RZ99,Long02}, in which usually only one aspect of the experiments has been considered.

An application of these findings to circumstellar disks using the Reynolds similarity principle can be found in Hersant et al.~\cite{HDH03} thereby providing a physical explanation of several observable indicators of turbulent transport.

\section{Taylor-Couette Flow}

\subsection{Stationary flow}
\label{statio}

The Taylor-Couette flow is obtained in the gap $d$ between two coaxial rotating cylinders of radii $r_{i,o}$, rotating at independent velocities $\Omega_{i,o}$. For the purpose of generality and to allow further comparison with astrophysical flows, the velocity field at the inner cylinder boundary may have a non-zero radial component.

The hydrodynamic equation of motions for an incompressible flow are given by:
\EQA
\partial_t {\bf u}+{\bf u}{\cdot}\nabla {\bf u}&=&-\frac{1}{\rho}\nabla p+\nu \Delta {\bf u},
\nonumber\\
\nabla\cdot {\bf u}&=&0.
\label{equans}
\ENA
where $\rho$ and $\nu$ are respectively the fluid density and kinematic viscosity, ${\bf u}$ is the velocity,  and  $p$ is the pressure.
Equation (\ref{equans}) admits a simple basic stationary solution, with axial and translation symmetry along the cylinders rotation axis (the velocity only depends on $r$). It is given by a flow with zero vertical velocity, and radial and azimuthal velocity given by:
\EQA
u_r&=&\frac{K}{r},\nonumber\\
u_\theta&=&A r^{1+\alpha}+\frac{B}{r},
\label{solutheta}
\ENA
where $A$ and $B$ are constants and $\alpha=K/\nu$. This basic laminar state depends on three constants $A$, $B$ and $K$, which can be related to the rotation velocities at the inner and outer boundaries\cite{Bahl70}:
\EQA
A&=&\frac{r_o^{-\alpha}}{1-\eta^{\alpha+2}}\left(\Omega_o-\eta^2\Omega_i\right),\nonumber\\
B&=&\frac{r_i^2}{1-\eta^{\alpha+2}}\left(\Omega_i-\Omega_o\eta^\alpha\right),
\label{CLGeneral}
\ENA
where $\eta=r_i/r_o$ and $\alpha=K/\nu=u_r(r_i)r_i/\nu$ is the radial Reynolds number, based on the radial velocity through the wall of the inner cylinder.

\begin{figure}[hhh]
\centering
\includegraphics[width=7cm]{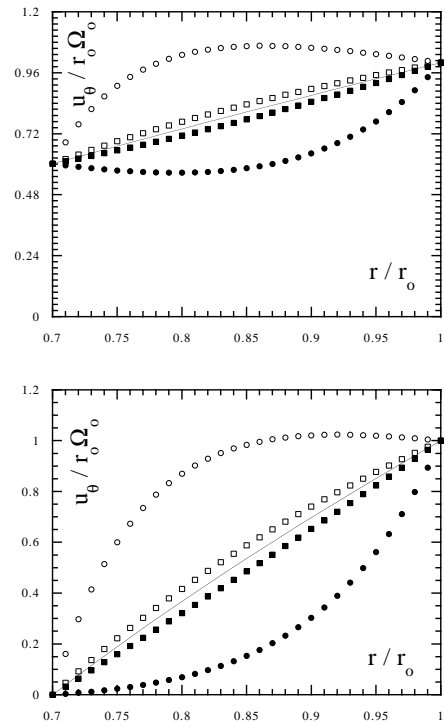}
\caption[]{Influence of the radial circulation onto the azimuthal profile. Line: case $\alpha=0$; $\square$: $\alpha=-1$; $\blacksquare$ $\alpha=1$; $\circ$: $\alpha=-10$; $\bullet$: $\alpha=10$. The upper panel is with $\Omega_i/\Omega_o=0.86$; the lower panel is with $\Omega_i/\Omega_o=0$. The radius ratio $r_i/r_o$ has been arbitrarily fixed at $0.7$}
\label{fig:profile-circ.eps}
\end{figure}

The radial circulation is quantified by the value of $\alpha$. It is positive for outward motions. For impermeable cylinders, $\alpha=K=0$ and one has the "classical" Taylor-Couette flows. For a porous internal cylinder, one obtains a Taylor-Couette flow with radial circulation.  The strength of the radial circulation can be controlled by using more or less porous cylinders~\cite{ML94}). Fig.~\ref {fig:profile-circ.eps} provides an example of the influence of the radial circulation on the azimuthal profile.

\noindent
In practice, even for impermeable cylinders, the flow is not purely azimuthal. Because of the finite vertical extent of the apparatus, a large-scale -- Ekman -- circulation is established through the effect of the top and bottom boundaries. This circulation depends on the ratio of radii and velocities, and on the top and bottom boundary conditions~\cite{Wen33,Rich01}. Its signature is easy to detect by profile monitoring, or by measuring the difference between the torque at the inner and outer cylinder~\cite{CvA66}. Of course this circulation is both radial and vertical and it varies along the cylinders axis. Also its intensity is not easy to control, since it is not fixed externally, but results from a non-trivial equilibrium within the flow. Still, at a given axial position, one may estimate this intensity by a fit of laminar profile using (\ref{solutheta}) and (\ref{CLGeneral}). To simplify the exploration of the parameter space, we shall restrict ourselves to the case of $\alpha\sim 0$, and study separately the influence of this parameter. In the laboratory, minimizing circulation effect is achieved by working with tall cylinders and consider only a fraction of the flow located at a distance to the top of about 1/3 of the total height, where the radial velocity is expected to be the weakest. Specific influence of $\alpha$ on stability and transport properties will be considered in Section \ref{stab-rad} and \ref{torque-rad}.

\subsection{Control parameters}
\label{contro}

Dimensional considerations show that there are only four independent non-dimensional numbers to characterize the system, which can be chosen in various ways.

\subsubsection{Traditional choice}
The traditional choice is to consider $d=r_o-r_i$ as the unit length, and $d^2/\nu$ as the unit time. With this choice, the dimensionless equations of motions are:
\EQA
\partial_t {\bf u^*}+{\bf u^*}{\cdot}\nabla {\bf u^*}
&=&-\nabla p^*+ \Delta {\bf u^*},
\nonumber\\
\nabla\cdot {\bf u^*}=0,
\label{equansadim1}
\ENA
with boundary conditions :
\EQA
{\bf u}^*(r_i)&=&(\alpha(1-\eta)/\eta,R_i,0)
\nonumber\\
{\bf u}^*(r_o)&=&(\alpha(1-\eta),R_o,0)
\label{CLadim1}
\ENA
where :
\EQA
\alpha&=&\frac{r_i u_r(r_i)}{\nu},\nonumber\\
\eta&=&\frac{r_i}{r_o},\nonumber\\
R_i&=&\frac{r_i\Omega_i d}{\nu},\nonumber\\
R_o&=&\frac{r_o\Omega_o d}{\nu}.
\label{param1}
\ENA
In the following, we will omit the star superscript indicating non-dimensional quantity. The present choice of unit amounts to define the control parameters by non-dimensional boundary conditions. When comparing flows that do not share identical geometry, it is of interest to identify control parameters characterizing the dynamical properties of the flows.

\subsubsection{Dynamics motivated choice}
In the case of rotating shear flows, it is convenient to write the equations in an arbitrary rotating frame with angular velocity $\Omega_{rf}$, choose $d$ as unit length, the inverse of a typical shear ${\tilde S}$ as unit time and $\tilde r$ as a typical radius. Furthermore, it is useful to introduce the "advection shear term" proposed in~\cite{Long02,PylOD04}
\begin{equation}\label{nabla'}
{\bm w}.{\bm\nabla}'{\bm w}\equiv{\bm w}.{\bm\nabla} (w_r {\bf
e}_r)+ [r{\bm w}.{\bm\nabla} (w_{\phi}/r)] {\bm e}_\phi + {\bm
w}.{\bm\nabla} (w_z {\bm e}_z).
\end{equation}
so that the contribution of the mean flow derivative to the modified advection term vanishes when the flow is not sheared, for azimuthal axisymmetric flow. As a result, one has
\EQA
\partial_t {\bf w}+{\bf w}{\cdot}\nabla'{\bf w}&=&-\nabla \pi - R_{\Omega}{\bf e_z}\times{\bf w}\nonumber\\
&&+R_{\cal C}\left(\frac{{w_{\phi}}^2}{r/{\tilde r}}{\bm e}_r -\frac{w_{\phi}w_r}{r/{\tilde r}}{\bm e}_{\phi}\right)\nonumber\\
&&+Re^{-1}\Delta {\bf w},\nonumber\\
\nabla\cdot {\bf w}&=&0,
\label{equansadim2}
\ENA
with boundary conditions :
\EQA
{\bf w}(r_{i,o})&=&{\bf u}(r_{i,o})-\frac{R_\Omega}{2 R_{\cal C}}\frac{r}{\tilde r}{\bf e_{\phi}}
\label{CLadim2}
\ENA
where :
\EQA
Re &=& \frac{{\tilde S} d^2}{\nu}\nonumber\\
R_\Omega &=& \frac{2{\Omega_{rf}}}{\tilde S}\\
R_{\cal C} &=& d/{\tilde r}
\label{param2}
\ENA
are the dynamical control parameters for a given radial circulation $\alpha$. $Re$ is an azimuthal Reynolds number, measuring the influence of shear. $R_\Omega$ is a rotation number, measuring the influence of rotation. Note that $\pi$ now also includes the centrifugal force term.

In this general formulation, one is free to choose $\Omega_{rf}$. It is convenient to choose $\Omega_{rf}$ as a typical rate of rotation $\tilde \Omega$ so that one can easily compare the Taylor-Couette case to the case of a plane shear in a rotating frame. For instance one can choose $\Omega_{rf}$ so that $w_{\theta}(r_i)=-w_{\theta}(r_o)$ in order to restore the symmetry between the two walls boundary conditions. This choice of $\tilde \Omega$ amounts to fix $\tilde r$ by $\Omega(\tilde r)=\tilde \Omega$. For consistency, it is then convenient to choose $\tilde S=S^{lam}(\tilde r)$. In this context and with $\alpha=0$, it is easy to relate the above control parameters $(Re,R_{\Omega},R_{\cal C})$ to the traditional choice $(R_i,R_o,\eta)$ :
\EQA
{\tilde r}&=&\sqrt{r_i r_o},\nonumber\\
Re&=&\frac{{\tilde S} d^2}{\nu}=
\frac{2}{1+\eta}\vert \eta R_o - R_i\vert,\nonumber\\
R_\Omega&=&\frac{2{\tilde\Omega}}{\tilde S}=
(1-\eta)\frac{R_i + R_o}{\eta R_o - R_i},\nonumber\\
R_{\cal C}&=&\frac{1-\eta}{\eta^{1/2}}.
\label{controlTC}
\ENA

The above control parameters have been introduced so that their definition apply to rotating shear flows in general and not only to the Taylor-Couette geometry. It is very easy in this formulation to relate the Taylor-Couette flow to the plane Couette flow with rotation, by simply considering the limit $R_{\cal C}\rightarrow 0$. Also, in the astrophysical context, one often considers asymptotic angular velocity profiles of the form $\Omega(r)\sim r^{-q}$ where $q$ then fully characterizes the flow. In that case $q=-\partial\ln\Omega/\partial\ln r=-2/R_{\Omega}$, which is a simple relation to situate astrophysical profiles in the control parameters space of the Taylor-Couette flows. From the hydrodynamic viewpoint, an important characteristic of the flow profile is the sign of the shear compared to the sign of the angular velocity, which defines cyclonic and anticylonic flows. For the co-rotating laminar Taylor-Couette flow, the sign of the local ratio $\Omega(r)/S(r)$ is constant across the whole flow and is thus simply given by the sign of the rotation number ($R_{\Omega}>0$ for cyclonic flows and $R_{\Omega}<0$ for anticylonic flows). Finally let us recall that an analogy exist between Taylor-Couette and Rayleigh-B\'enard convection (see ~\cite{DubrHers02} for details and ~\cite{Prig04} for a review), which calls for an even larger generalization of the control parameters definition.

Figure~\ref{fig:param} displays the characteristic values taken by the new parameters in the usual parameter space $(R_i,R_o)$ for co-rotating cylinders. It also helps to situate cyclonic and anti-cyclonic flows, as well as prototypes of astrophysical flows.

\section{Stability properties}
\label{stab} 

\subsection{Inviscid limit and data sources for viscid flows}
\label{stab-invi}

As usual when considering stability properties, one must distinguish stability against infinitesimal disturbances -- linear stability -- from that against finite amplitude ones -- non-linear stability. When the basic flow is unstable against finite amplitude disturbance, but linearly stable, it is called subcritical, by contrast with the supercritical case for which the first possible destabilization is linear (see ~\cite{Jo76,DauMan} for further details).

In the inviscid limit ($Re\rightarrow\infty$), and for axisymmetric disturbances, the linear stability properties of the flow are governed by the Rayleigh criterion. The fluid is stable if the Rayleigh discriminant is everywhere positive:
\begin{equation}
\frac{\Omega}{r} \partial_r L(r)>0,
\label{Rayleigh}
\end{equation}
where $L(r)=r^2\Omega(r)$ is the specific angular momentum. Applying this criterion to the laminar profile leads to
\begin{equation}
(R_\Omega+1)(R_\Omega+1-{\tilde r}^2/r^2)>0.
\label{newrayleigh}
\end{equation}

\begin{figure}[hhh]
\centering
\hspace{-1cm}
\includegraphics[width=8cm, height=7cm]{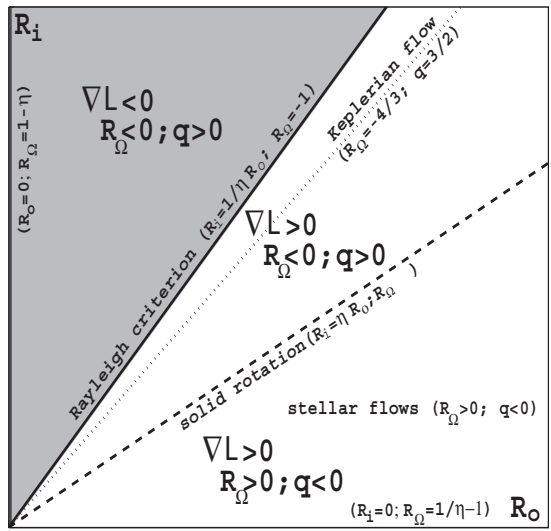}
\caption{Parameters space and some TC flow properties for co-rotating cylinders ($\eta=0.72$). Flows with positive gradient of angular momentum $L=r^2\Omega$ but negative (resp. positive) gradient of angular velocity are referred to as Keplerian (resp. stellar). The shaded area corresponds to Rayleigh unstable flows (supercritical case).}
\label{fig:param}
\end{figure}

\noindent
Since ${\tilde r}/r$ varies between $1/\eta$ and $\eta$, one obtains that in the inviscid limit, the flow is unstable against infinitesimal axisymmetric disturbances when ${R_{\Omega}^{\infty}}^{-}<R_{\Omega}<{R_{\Omega}^{\infty}}^{+}$, where ${R_{\Omega}^{\infty}}^{-}=-1$, respectively ${R_{\Omega}^{\infty}}^{+}=1/\eta-1$, are the marginal stability thresholds in the inviscid limit (superscript $\infty$) in the cyclonic case ($R_\Omega>0$, subscript $+$), respectively anticyclonic case ($R_\Omega<0$, subscript $-$). These Rayleigh limits are also displayed on figure~\ref{fig:param}, where they have to be seen as asymptotic. As a matter of fact, this information is rather poor : 

\begin{itemize}
\item non-axisymmetric disturbances can be more destabilizing than axisymetric ones, so that the flow could be linearly unstable in part of the linearly stable domain;
\item viscous damping will probably reduce the linearly unstable domain; 
\item finally, finite amplitude disturbances may seriously reduce the stable domain.  
\end{itemize}

In the following, assuming that the axisymmetric disturbances are indeed the most dangerous one at the linear level -- which up to now is validated both experimentally and numerically --, we will consider the two last items.
On one side, we will review the existing results on the effect of viscosity in the supercritical case, which will provide us with a critical Reynolds number as a function of the other parameters $R_c(R_\Omega,\eta)$. On the other side, we will investigate the subcritical stability limit, when the flow is linearly stable and try to figure out what is the behavior of the minimal Reynolds number for self-sustained turbulence $R_g(R_\Omega,\eta)$.
 
These boundaries can be estimated via different tools, depending of the type of experiment and available measurements. In numerical experiments, the simplest way to estimate the stability boundary in the linear case is through a modal decomposition and a monitoring of real part of the largest eigenvalue. In laboratory experiments, at least three different tools have been used: i) torque measurements; ii) flow visualization; iii) mean velocity profile measurements. Torque measurements have been traditionally used in the past~\cite{Tay36,Wen33}. Their advantage is their accuracy and their flexibility to detect other transition at larger Reynolds numbers. Their inconvenience is their difficulty of implementation in the case where both cylinders are rotating. Flow visualizations allow discriminating between laminar and turbulent flows but suffer from the lack of quantitative information on the flow. Mean velocity profile measurement is a third alternative, which allows determination of critical Reynolds number from deviation of velocity profiles with respect to laminar value, or changes of regime. This technique is more local in nature, and requires advanced techniques of in-flow measurements. In the sequel, we shall use data from several sources, described in table~\ref{tab_source}. Except for the data of Richard, all of them have been published. Those by Richard are available in his thesis manuscript~\cite{Rich01}. We take the opportunity of this synthesis to integrate them in a larger perspective.

\begin{table}[h]
\begin{center}
\begin{tabular}{lll}
$\eta$ & $R_{\Omega}$ & source \\
\hline
1 & $[0, 0.1]$ & Tillmark et al.~\cite{TilmAlfr96} \\
0.983 & $\simeq 0$ & Prigent et al.~\cite{PGCDS02} \\
0.724 & $-0.276$ & Lewis et al.~\cite{LewiSwin99} \\
0.7 & $[0, 0.6]$ & Richard~\cite{Rich01} \\
0.7 & $[-1.5, -1]$ & Richard~\cite{Rich01} \\
$[0.79, 0.97]$ & $1/\eta -1$ & Taylor~\cite{Tay36} \\
$0.68, 0.85, 0.93$ & $[-0.7, 0.5] $ & Wendt~\cite{Wen33} \\
\end{tabular}
\caption{Experimental data and sources}
\label{tab_source}
\end{center}
\end{table}

\subsection{Super-critical case}
\label{stab-super}

\begin{figure}[hhh]
\centering
\includegraphics[width=8cm]{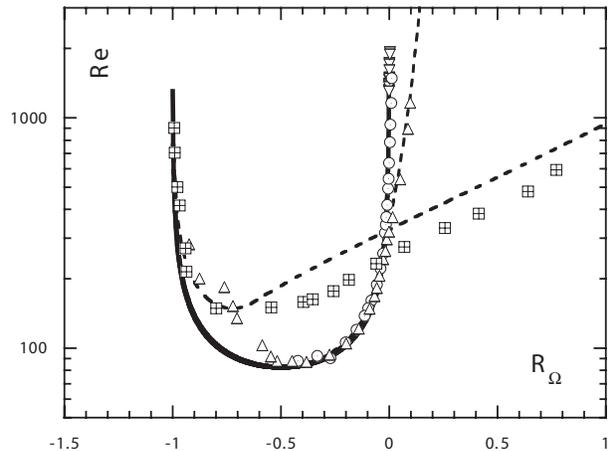}
\caption[]{The linear stability boundary. Numerical data by Snyder $\boxplus$: $\eta=0.2$; $\triangle$: $\eta=0.8$, $\odot$ $\eta=0.935$. Experimental data by Prigent et al. $\nabla$: $\eta=0.983$. Continuous line: Lezius and Johnston plane Couette or small gap limit stability criteria. Dashed line : Esser and Grossman prediction for $\eta=0.8$ and $\eta=0.2$.}
\label{fig:Rc}
\end{figure}

Numerous experimental set ups were used to study the stability boundary in the linear case, starting from the early experiments of Couette~\cite{Coue90}, Taylor~\cite{Tayl23}, and Donnelly and Fultz~\cite{DonnFult60}. The viscosity damps the instability until $Re>R_c(R_\Omega,\eta)$, corresponding to the  transition from the laminar flow to the so-called Taylor vortices flow.
Figure~\ref{fig:Rc} displays the numerical data by Snyder~\cite{Sny68}, providing the stability threshold $R_c$ as a function of $R_\Omega$ for three gap size ($\eta=0.935, 0.8, 0.2$), and illustrates the influence of the curvature on the instability threshold. The experimental data of Prigent et al.~\cite{PGCDS02}, at $\eta=0.983$ is also reported.

As $\eta\to 1$ (rotating plane Couette limit), the stability curve becomes symmetric around $R_\Omega=-1/2$ and diverges at $R_\Omega=0$ or $-1$. This is in agreement with the linear stability criterion for the rotating plane Couette flow~\cite{Brad69,Ped69,LeziJohn76}, a generalization of the first exact result giving the linear stability of the non-rotating plane Couette flow for all Reynolds number~\cite{Rom73}. The observed symmetry actually reflects symmetry in the rotating plane Couette. The linearized equations of motions are invariant by the transformation exchanging streamwise and normal to the walls coordinates and velocities (corresponding to exchanging $r$ with $\phi$ and $u_r$ and $u_\phi$, in Taylor-Couette). This transformation changes $R_\Omega$ into $-1-R_\Omega$, hence the symmetry around $R_\Omega=-1/2$. When $\eta$ becomes smaller than $1$, curvature enters into play and breaks the symmetry resulting in less and less symmetrical curves, as can be observed for $\eta=0.2$. 

\begin{figure}[hhh]
\centering
\includegraphics[width=7.5cm, height=7cm]{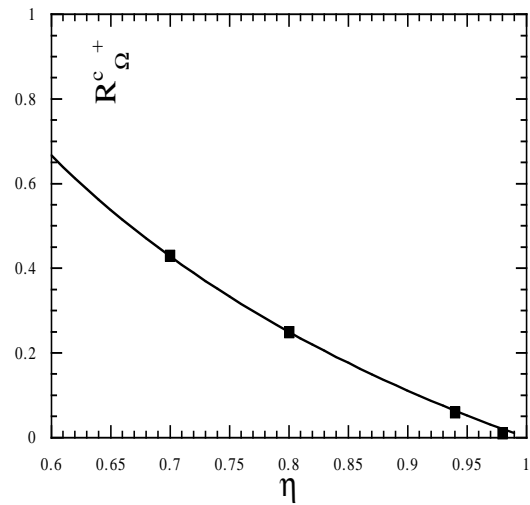}
\caption{The critical rotation number $R_\Omega^{c+}$ as a function of the gap size $\eta$. $\blacksquare$: Computed from the analytic formula of Esser and Grossman. Plain line: $(1-\eta)/\eta$.}
\label{fig:romc}
\end{figure}

The above stability boundary can be recovered numerically by classical stability analysis, using e.g. normal mode analysis with numerical solutions~\cite{Chan60}. Interestingly a very good approximate analytical formula in the whole parameter space has recently been derived by Esser and Grossmann~\cite{EG96}. It is:
\EQA
R_{c}^2\left(R_\Omega+1\right)\left(R_\Omega+1-\frac{1}{\eta x^2}\right)=&&\nonumber\\
-1708\left(\frac{(1-\eta)}{2\eta(x(\eta)-1)}\right)^4,&&
\label{Getransforme}
\ENA
with
\EQA
&&x(\eta)=1+\frac{1-\eta}{2\eta}\Delta\left(a(\eta) \frac{d_n}{d}\right),\nonumber
\ENA
\EQA
&&\frac{d_n}{d}=\frac{\eta}{1-\eta}\left(\frac{1}{\sqrt{\eta(R_\Omega+1)}}-1\right),\nonumber\\
&&a(\eta)=(1-\eta)\left(\sqrt{\frac{(1+\eta)^3}{2(1+3\eta)}}-\eta\right)^{-1},
\label{Getransforme2}
\ENA 
where $\Delta(x)$ is a function equal to $x$ if $x<1$ and equal to $1$ if $x>1$. Continuous lines on figure~\ref{fig:Rc} give a good insight on the validity of the above formula. Let us underline that in the absence of the above formula, small and wide gap approximations were often used. Whereas the small gap approximation 
\begin{equation}
R_{c}^{sg}=\sqrt{\frac{1708}{-R_\Omega(R_\Omega+1)}}.
\label{sglimit}
\end{equation}
works rather well until $\eta=0.8$, the large gap one
\begin{equation}
\eta R_{c}^{wg}=\sqrt{\frac{1708}{4}}\frac{1}{\Theta(R_\Omega+1)},
\label{wglimit}
\end{equation}
where $\Theta(x)$ is the Heavyside function, remains a very poor approximation even for $\eta=0.2$. Note that the formula (\ref{Getransforme}) defines two critical rotation number for which the critical Reynolds number diverges: $R_\Omega^{c-}=-1$ and $R_\Omega^{c+}$ such that $R_\Omega^{c+}=-1+1/\eta x^2$. This number has been computed for various $0.7<\eta<1$ and is shown on figure \ref{fig:romc}. One sees that it is very well approximated by the formula $R_\Omega^{c+}=(1-\eta)/\eta$. This remark is used in the next section.\

In this supercritical situation, the flow undergoes several other bifurcations following the first linear instability and turns into more and more complex patterns, eventually leading to turbulence.
Interestingly, at much larger Reynolds number, an additional transition have been reported~\cite{LFS92}. One indeed observes a change in the torque dependence on the Reynolds number, which could be associated with a featureless turbulence regime. Sometime called "hard turbulence", this regime is observed for $Re>R_T$. For reasons that will become clearer, we defer its discussion after study of the torque.

\subsection{Sub-critical case}
\label{stab-sub}

\begin{figure}[hhh]
\centering
\includegraphics[width=7.5cm, height=7cm]{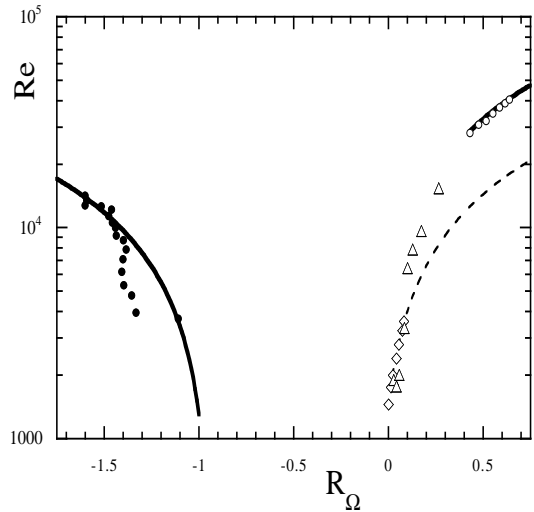}
\caption{The non-linear stability boundary. Cyclonic flow: $\triangle$: Taylor data with inner cylinder at rest and $0.7<\eta<0.935$; $\circ$: Richard data with $\eta=0.7$; $\lozenge$: Tilmark data for rotating plane Couette flow $\eta=1$; dotted line: linear fit of Tilmark's data; plain line: linear fit of Richard's data. Anticyclonic flow: $\bullet$: Richard data with $\eta=0.7$; plain line: linear fit of Richard's data.}
\label{fig:nonlin}
\end{figure}

In the absence of general theory for globally subcritical transition to turbulence, the non-linear stability boundary has only been explored experimentally. Wendt~\cite{Wen33} and Taylor~\cite{Tay36} consider the case with inner cylinder at rest, corresponding to $R_\Omega=1/\eta-1>0$, at various gap size, using torque measurements. A more recent experiment by Richard~\cite{Rich01} explores the domain $-1.5<R_\Omega<-1$ and $R_\Omega>0.5$, at fixed gap size $\eta=0.7$, using flow visualizations. Finally one also has the measurement conducted in a rotating plane Couette flow ($\eta=1$) by Tillmark and Alfredsson~\cite{TilmAlfr96} for $R_\Omega>0$. The corresponding results are reported on figure~\ref{fig:nonlin}, giving $R_{g}$ as a function of $R_\Omega$ for different value of $\eta$.

One must be very cautious when looking at this naive representation of the data, especially on the cyclonic side. First, the data are presented for different values of $\eta$. Especially for the data of Taylor ($\triangle$), each point is a different $\eta$. The fact that the data look aligned through all values of $\eta$ is an artifact of the representation as illustrated by the extrapolation of the linear fit of Tillmark's data. Second, looking at this figure $R_\Omega=0$ seems to play a similar role in the cyclonic regime than $R_\Omega=-1$ in the anticyclonic case. As we have seen above, when studying the linear stability, this is true for $\eta=1$ only. As discussed in previous section, the correct value of the marginal stability is approximately equal to the inviscid limit for the cyclonic case $R_\Omega^{c+}\approx {R_\Omega^{\infty }}^{+}=1/\eta-1$. Taylor's data are actually given at this precise value of $R_\Omega$, because Taylor performed his experiments with the internal cylinder at rest. This condition imposes $R_\Omega=1/\eta-1$, which {\it coincides} with the marginal stability limit. In the following, we shall try to extract from this data the maximal knowledge about the dependence of ${R_g}^{(+,-)}$ on $R_\Omega$ and $\eta$, both in the cyclonic and anticyclonic case.

All the data about the manifold ${R_g}^{(+,-)}(R_\Omega,\eta)$ are obtained close to its intersection with the manifold $R_\Omega={R_\Omega^{c}}^{(+,-)}$. Therefore one first has to estimate the locus of the intersection between these two manifolds that is ${R_g}^{(+,-)}({R_\Omega^{c}}^{(+,-)}(\eta),\eta)=f^{(+,-)}(\eta)$, then the variation of ${R_g}^+$ with $R_\Omega$, close to the manifold, at the intersection.

\begin{figure}[hhh]
\centering
\includegraphics[width=7.5cm]{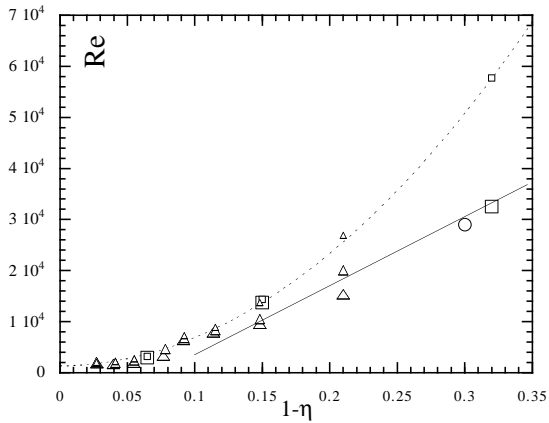}
\caption{Subcritical thresholds in the cyclonic regime, obtained with the inner cylinder at rest, that is $R_\Omega={R_\Omega}^{c +}$. $\square$: Wendt's data; $\triangle$: Taylor's data; $\circ $: Richard's data; plain and dotted line: fit of $f^{+}(\eta)={R_g}^+({R_\Omega^{c}}^{+}(\eta),\eta)$ (see text for details). The size of the symbol denotes different estimate by Richard and Zahn (small), Zeldovich (medium) and present authors (large) based on published figures of Taylor and Wendt.}
\label{fig:f_eta}
\end{figure}

Let us first consider the cyclonic case. One can take benefit of Taylor's and Wendt's data to estimate $f^{+}(\eta)$ as proposed by Richard and Zahn~\cite{RZ99}. The fact that the data are read from the original figure of Taylor and Wendt however induces a natural error bar in the determination of the critical Reynolds number, as illustrated on figure~\ref{fig:f_eta}, where several estimates, obtained by different authors,  are reported.

Because of this error, it is difficult to give a precise fit of the function $f^+$. One sees that the quadratic regime in $1-\eta$ given by $f^{+}(\eta)=1400+550000(1-\eta)^2$ and proposed by Richard and Zahn provides a good upper estimate of the function. A linear trend in $1-\eta$, with slope 
$136000$ gives a good lower estimate of the data for $1-\eta<0.1$, as shown on figure~\ref{fig:f_eta}. Clearly, more precise estimate of this function using modern data will be welcome. Note that at $\eta\to 1$, 
the function tends to a constant $f^{+}(1)=1400$ that is nothing but $R_{g}^{PC}=1400$, the global stability threshold measured independently by Tillmark and Alfredsson~\cite{TilmAlfr96}  and Dauchot and Daviaud~\cite{DauDav94} in the non-rotating plane Couette flow.
The second step is to propose a linear development in $R_\Omega-{R_\Omega^{c }}^{+}$, close to the above estimate:
\begin{equation}
R_{g}^{+}(R_\Omega,\eta)=f^{+}(\eta)+a^{+}(\eta)\left(R_\Omega-{R_\Omega^{c }}^{+}(\eta)\right).
\label{Gene_cyc_sub}
\end{equation}
For $\eta=1$ one recovers the linear fit proposed by Tillmark and Alfredson (plotted and extrapolated on fig~\ref{fig:nonlin}) for the rotating plane Couette flow:
\begin{equation}
R_{g}^{+}(R_\Omega,\eta=1)=1400+26000 R_\Omega.
\label{Till_cyc_sub}
\end{equation}
that is $a^{+}(\eta=1)=26000$. For $\eta=0.7$, the linear fit of Richard's data (plotted and extrapolated on fig~\ref{fig:nonlin}) leads to $a^{+}(\eta=0.7)=59000$.

In the anticyclonic case, the situation is simpler because ${R_\Omega^c}^{-}=-1$, does not depend on $\eta$. On the other hand, data are available for a unique value $\eta=0.7$, so that one cannot estimate $f^{-}(\eta)$. The only fit that can be performed in this state of experimental knowledge is:
\begin{equation}
R_{g}^{-}(R_\Omega,0.7)=f^{-}(0.7)+a^{-}(0.7)\mid R_\Omega-{R_\Omega^c}^{-}\mid.
\label{Gene_cyc_sup}
\end{equation}
One finds $f^{-}(0.7)=1300$ and $a^{-}(0.7)=21000$ and the fit is displayed on figure~\ref{fig:nonlin}. In the anticyclonic regime, at least for this value of $\eta$, one recovers a dependence on the rotation similar to that of the plane Couette flow. Also remarkable is the fact that $f^{-}$ is so close to $R_{g}^{PC}$ in the non-rotating case.

Altogether the data collected to date suggest that, in the linearly stable regime, the Reynolds number of transition to subcritical turbulence be well represented by
\begin{equation}\label{Rgpm}
R_g^{\pm}(R_\Omega, \eta)=f^{\pm}(\eta) + a^{\pm}(\eta)|R_\Omega - R_\Omega^{c,\pm}|,
\end{equation}
with $1\times 10^5(1-\eta)^{1}<f^{+}(\eta)<1400+5.5\times 10^5(1-\eta)^{2}$,$f^{-}(0.7)=1300$ and $21000 \lesssim a^\pm
\lesssim 59000$. It is difficult to distinguish the effect of experimental procedures from the effects of gap width dependence in the present parameter range.

\subsection{Influence of radial circulation}
\label{stab-rad}

\subsubsection{Super-critical case}
\begin{figure}[hbt]
\centering
\includegraphics[width=7.5cm]{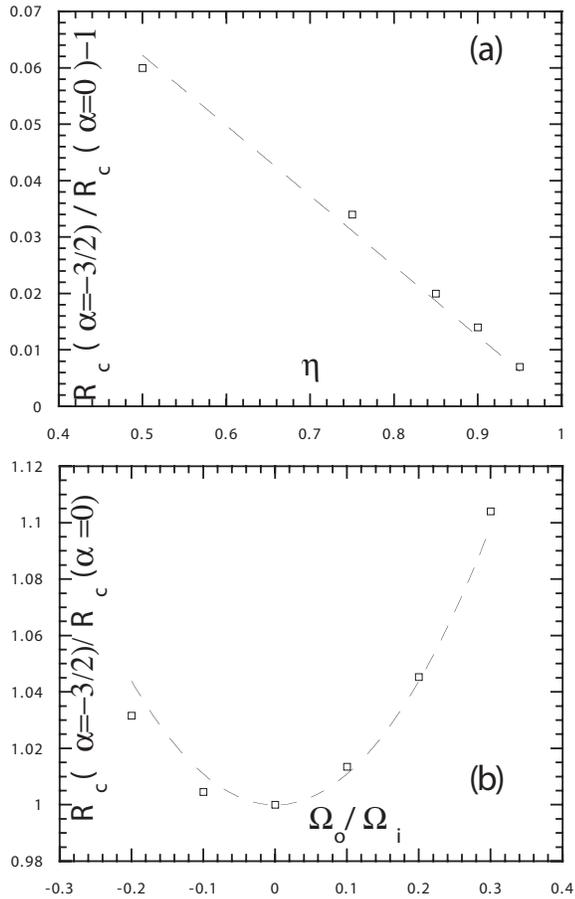}
\caption[]{$R_{c}(\alpha=-3/2)/R_{c}(\alpha=0) -1$ as a function of  (a) $\eta$ for $\Omega_0/\Omega_i=0$; (b) $\Omega_0/\Omega_i=0$ for $\eta=0.85$. $\square$ : data from Min and Lueptow. The dotted lines are the fit eq. (\protect\ref{fitML1}) and (\protect\ref{fitML2}).}
\label{fig:rad}
\end{figure}

The influence of radial circulation on the linear stability onset has been studied numerically by Min and Lueptow~\cite{ML94}. They observed that an inward radial flow and strong outward flow have a stabilizing effect, while a weak outward flow has a destabilizing effect. We may use their data to get more precise estimates for the case $\alpha=-3/2$, ($q=3/2$, Keplerian case). Figure~\ref{fig:rad} shows the ratio $R_c(\alpha=-3/2)/R_c(\alpha=0)-1$ as a function of $\eta$ for $\Omega_o/\Omega_i=0$. One sees that the variation is quasi-linear.

\noindent
A best fit gives:
\begin{equation}
\frac{R_{c}(\alpha=-3/2)}{R_{c}(\alpha=0)}=1+0.12(1-\eta), \quad\frac{\Omega_o}{\Omega_i}=0.
\label{fitML1}
\end{equation}
On the same graph, we show $R_{c}(\alpha=-3/2)/Re_{c}(\alpha=0)-1$ as a function of $\Omega_o/\Omega_i$ for $\eta=0.85$. A best fit gives:
\begin{equation}
\frac{Re_{+}(\alpha=-3/2)}{Re_{+}(\alpha=0)}=1+1.16\left(\frac{\Omega_o}{\Omega_i}\right)^2,\quad \eta=0.85.
\label{fitML2}
\end{equation}
 
\subsubsection{Sub-critical case}
The influence of the radial circulation on the non-linear stability has not been systematically studied. However, we can get partial answers from the experiments of Wendt~\cite{Wen33} and Richard~\cite{Rich01}, where the influence of the top and bottom circulation on the onset of stability has been studied. Both Richard and Wendt investigated the stability boundary with different boundary conditions. One boundary condition was with the bottom attached to the outer cylinder. In this case, the circulation is mainly in the anti-clockwise direction, with radial velocities outwards at the bottom ($\alpha>0$). Another boundary condition was with the bottom attached to the inner cylinder (at rest). In that case, the circulation is in the opposite direction, with inward radial velocities at the bottom ($\alpha<0$). A last boundary condition was intermediate between the two, with only part of the bottom attached to the outer cylinder. In neither case, noticeable change of the stability boundary has been noticed, which means that at this aspect ratio $\eta=0.7$, the radial circulation induced by the boundary conditions has an impact on the subcritical threshold Reynolds number which is less than 10 per cent (accuracy of the measurements).

\subsection{Influence of aspect ratio}
\label{stab-ratio}

\begin{figure}[hhh]
\centering
\includegraphics[width=7.5cm]{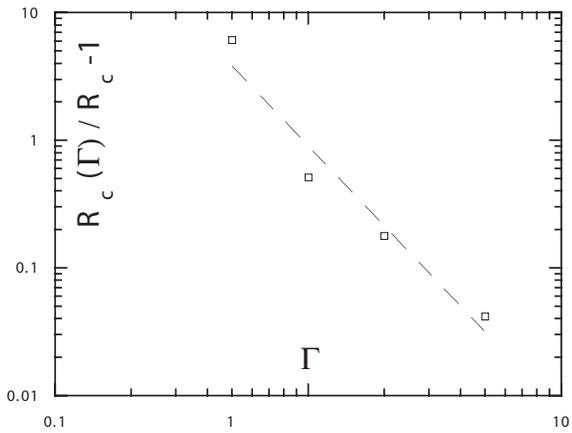}
\caption[]{Influence of aspect ratio onto the critical Reynolds number for instability, in the case where the outer cylinder is at rest. $\square$: numerical data by Chandrasekhar. The dotted line is a power law fit $0.9 \Gamma^{-2}$.}
\label{fig:aspect}
\end{figure}

Most of the experimental set-ups described in this paper have a very large aspect ratio $\Gamma=H/d\gg 1$. Keplerian disks are characterized by a small aspect ratio $H/d\sim 0.01-0.1$. It would be interesting to conduct systematic studies of the variation of $\Gamma$ onto the stability and transport properties. The influence of $\Gamma$ onto the instability threshold, in the case of outer cylinder at rest has been computed by Chandrasekhar~\cite{Chan60}, Snyder~\cite{Sny68}. This is illustrated in Fig. \ref{fig:aspect}. The critical Reynolds number is increased, as $\Gamma$ is decreased. It follows an approximate law:
\begin{equation}
R_{c}(\Gamma)=R_{c}\left(1+\Gamma^{-2}\right)+ O(\Gamma^{-2}).
\label{aspectloi}
\end{equation}
This behavior can be understood if one says that as $\Gamma$ becomes smaller, the smallest relevant length scale in the problem become $H$ instead of $d$. The relevant Reynolds number has thus to be corrected by a factor $(H/d)^2$, hence, the $\Gamma^{-2}$ law. However, another experimental study by Park et al.~\cite{PCD81} suggests that the physical relevant length scale is $\sqrt{Hd}$ instead of $H$. A possible explanation of the difference is through the Ekman circulation, which is present in experiments and not in numeric. This circulation may couple vertical and radial velocities, leading to an effective length scale. The only way to settle this issue is through smaller aspect ratio systematic laboratory and numerical experiments.

\subsection{Structural stability}
\label{stab-struc}

To close this section, it is interesting to consider the influence of additional physical forces that may be relevant to astrophysical flows. In the sequel, we only give a summary of the main experimental or theoretical results obtained, referring to the publications for more details.

\subsubsection{Magnetic field}
\label{stab-struc-mag}

The influence of a vertical magnetic field on the stability of a Taylor-Couette flow has been studied theoretically~\cite{Veli59,Chan60} and experimentally by Donnelly and Ozima~\cite{DonnOzim62} using mercury. Applications to astrophysics have been discussed by Balbus and Hawley~\cite{BH91}. This motivated a lot of numerical work on this instability. For references, see e.g. ~\cite{RSS03}.\

In the inviscid limit, the presence of a magnetic field changes the Rayleigh criteria (\ref{Rayleigh}). For example, in the case of a magnetic field given by $B=\sqrt{\mu_0\rho}(0,H_\theta(r), H_z)$, the sufficient condition for stability is now~\cite{HowaGupt62,DubrKnob92}:
\begin{equation}
r^2\partial_r\Omega^2-\frac{1}{r^2}\partial_r \left(r^2 H_\phi^2\right)>0.
\label{magnetoRayl}
\end{equation}
Therefore, anti-cyclonic flow, with $R_\Omega<0$ are now potentially linearly {\sl unstable} in the presence of a magnetic field with no azimuthal and radial component~\cite{BH91}.\

\begin{figure}[hhh]
\centering
\includegraphics[width=7.5cm]{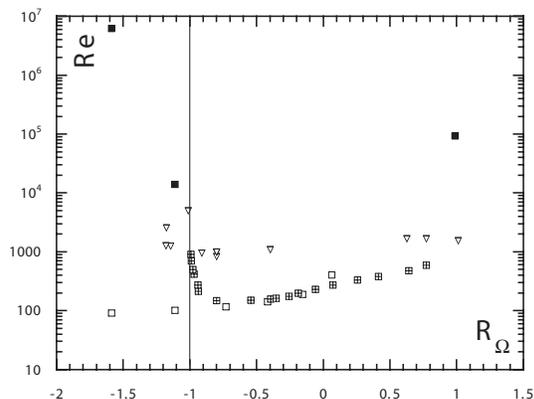}
\caption[]{Influence of body forces on the linear stability boundary. $\boxplus$: without forces, $\eta=0.2$, numerical data from Snyder (1968). With vertical constant magnetic field, at $Pm=1$ ($\square$) and $Pm=10^{-5}$ ($\blacksquare$), $\eta=0.27$; numerical data from R\"udiger et al, 2003; $\nabla$: with vertical stratification, $\eta=0.2$; data from Whithjack and Chen (1974).}
\label{fig:lin-mag}
\end{figure}

The linear instability in the presence of dissipation has only been studied numerically. A first observation was that boundary conditions (e.g. insulating or conducting walls) are relevant to determining the asymptotic behaviors~\cite{Chan60}. The proposed explanation is that the magnetic field makes the flow adjoin the walls for longer distances, so that the viscous dissipation remains comparable to the Joule dissipation at all fields. A second observation is the importance of the magnetic Prandtl number $P_m=\nu/\kappa_m$ ($\kappa_m$ is the magnetic diffusivity) on the instability~\cite{WillBar02,RSS03}. On general grounds, it seems that at small Prandtl numbers, the magnetic field {\sl stabilizes} the flow in the supercritical case, while at large Prandtl numbers, the magnetic field {\sl destabilizes} the flow. In the subcritical case, the magnetic field can excite a linear instability for anti-cyclonic flow, at any Prandtl number. This is illustrated in Fig. \ref{fig:lin-mag}.\

Scaling of critical Reynolds number with magnetic Prandtl numbers have been found: in the supercritical case $-1<R_\Omega<0$, the critical Reynolds number scales like $P_m^{-1/2}$~\cite{WillBar02}. In the subcritical case $R_\Omega=-1$, the critical Reynolds number scales like $Pm^{-1}$. 

\subsubsection{Vertical stratification}
\label{stab-struc-strat}

A vertical stable stratification added onto the flow plays the same role as a vertical magnetic field at low $Pm$. In the inviscid limit, its presence changes the Rayleigh criteria into $r^2\partial_r\Omega^2>0$~\cite{MWY01,Dubr03}) . This means that all anti-cyclonic flows are potentially linearly unstable. The role of dissipation on the instability has been studied numerically~\cite{YWM01,Dubr03}) and experimentally~\cite{WithChen74,BoubHopf97}. It was found that stratification stabilizes the flow in the GSPC regime, while it destabilizes it in the GSBC anti-cyclonic regime. The critical Reynolds number was found to scale with the Froude number (ratio of rotation frequency to Brunt Vaissala frequency) like $Fr^{-2}$, and to scale with the Prandtl number (ratio of viscosity to heat diffusivity) like $Pr^{-1/2}$.

\subsubsection{Radial stratification}
\label{stab-struc-rad}

A radial temperature gradient applied to the flow changes the stability. In the inviscid limit, the Rayleigh criterion is modified by the radial temperature gradient into~\cite{ChenKuo90,MGD01}:
\begin{equation}
\frac{\Omega}{r} \partial_r (r^2\Omega)\left(1-\frac{\Delta T}{T_0}\right)-\frac{\partial_r T}{T_0}r\Omega^2>0,
\label{Rayleighradial}
\end{equation}
where $1/T_0$ is the coefficient of thermal expansion and $\Delta T$ is the temperature difference between the cylinders. The last term in (\ref{Rayleighradial}) induces an asymmetry between the case with positive $\Delta T$ and negative $\Delta T$. An experimental study by Snyder and Karlsson~\cite{SnyKarl64} helps to quantifying the role of dissipative processes. It was found that both positive and negative $\Delta T$ have a stabilizing effect when $\Delta T$ is small, and a destabilizing effect when $\Delta T$ is large. A more complete exploration of the parameter space would be welcome, since astrophysical disks are likely to be subject to this kind of stratification. 

\subsubsection{Summary}
These studies point out an interesting dissymmetry between the case $R_\Omega>0$ (cyclonic flows) and $R_\Omega<0$ (anti-cyclonic flows). In many instances, the regime of linear instability is {\sl extended} by the large scale force into the whole domain $R_\Omega<0$. As a result, in the anticyclonic regime one often has to deal with a competition between a linear destabilization mechanism induced by the large scale effect and the subcritical transition controlled by the self-sustained mechanism of the turbulent state.

\section{Mean flow profiles}

\subsection{Supercritical case}
\label{mean-super}

Turbulent mean profiles have been measured recently for different Reynolds number by Lewis and Swinney~\cite{LewiSwin99} in the case with outer cylinder at rest. They observe that the mean angular momentum $L=r{\bar u}_\theta$ is approximately constant within the core of the flow: $L\sim 0.5 r_i^2\Omega_i$ for Reynolds numbers between $1.4\times 10^4$ and $6\times 10^5$. At low Reynolds number, this feature can be explained by noting that reducing the angular momentum is a way to damp the linear instability, and, thus, to saturate turbulence. At larger Reynolds number, however, one expects the turbulence to be sustained by the shear in the same way as it is when there is no linear instability at all. Accordingly, this constancy of the angular momentum is quite a puzzling fact. Some understanding of this behavior can be obtained by observing that the mean profiles obtained by Lewis and Swinney are actually in good agreement with a profile obtained by Busse upon maximizing turbulent transport in the limit of high Reynolds number\cite{Busse72,Busse96}:
\begin{equation}
u_\theta^\infty(r)=-\eta \frac{{\tilde r}^2 {\tilde S}}{8 r}+r\left({\tilde \Omega}+\frac{2-3\eta+2\eta^2}{4(1+\eta^2)}{\tilde S}\right).
\label{Busse}
\end{equation}
This profile bears some analogy with the laminar profile, which reads:
\begin{equation}
u_\theta^{lam}(r)=-\eta \frac{{\tilde r}^2 {\tilde S}}{2 r}+r\left({\tilde \Omega}+\frac{1}{2}{\tilde S}\right).
\label{laminar}
\end{equation}
In the Busse solution, the shear profile $S^\infty(r)=\eta {\tilde r}^2 {\tilde S}/4r^2= 1/4 S^{lam}(r)$. This ratio is analog to the value observed at very large Reynolds number in the non-rotating plane Couette flow~\cite{Busse70}. It is therefore a clear signature of the shear instability, with no discernable influence of rotation, at least for the limited value of  the rotation number (of the order of $-0.28$) considered by Lewis and Swinney. So it is interesting to test the Busse asymptotic profile using other data, with different rotation number. This will be the purpose of the next section, where Richard data will be used.

We may however not conclude this section without noting an intriguing property of the Busse solution. Considering ${R_\Omega}^{turb}=2{\Omega}^\infty(\tilde r)/S^{\infty}(\tilde r)$, we get from (\ref{Busse}):
\begin{equation}
{R_\Omega}^{turb}=4R_\Omega+3\frac{(1-\eta)^2}{(1+\eta^2)}.
\label{Busserotnumb}
\end{equation}
So the condition $-1<{R_\Omega}^{turb}<1/\eta-1$ ("linear stability of the turbulent profile") is satisfied provided $R_\Omega$ follows:
\begin{equation}
-\frac{2-3\eta+2\eta^2}{2(1+\eta^2)}<R_\Omega<\frac{1-\eta}{4\eta}\frac{(1+\eta)^2-3\eta(1-\eta)^2}{(1+\eta^2)},
\label{Bussecond}
\end{equation}
that is, in the small gap limit, $-1/4<R_\Omega<0$. As we shall see in the sequel, this is precisely the range of value where the torque is extremum.

\subsection{Subcritical case}
\label{mean-sub}

\begin{figure}[h!]
\centering
\includegraphics[width=7.5cm]{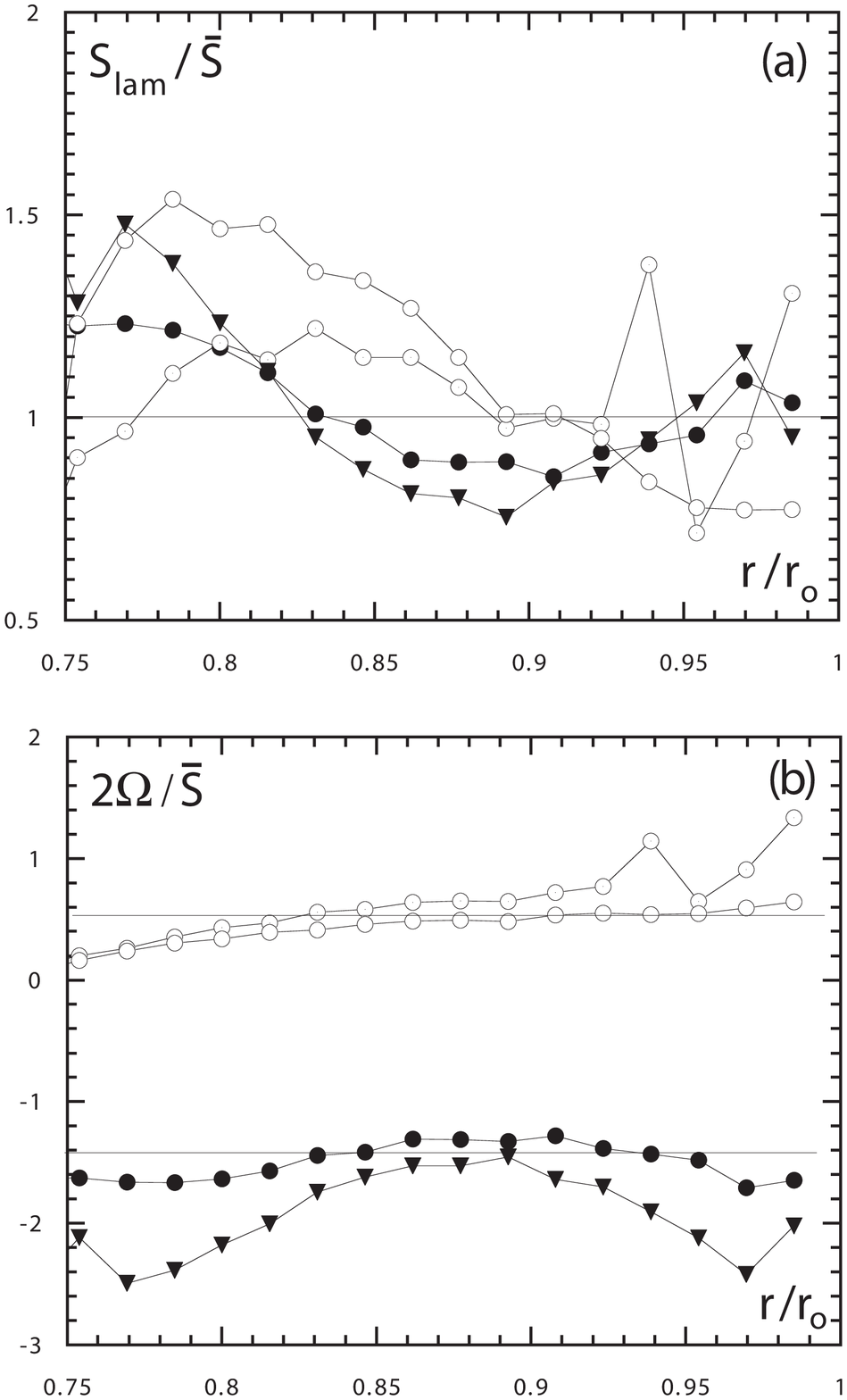}
\caption[]{Mean velocity profiles from Richard at $Re/R_g=1.6$: (a): cyclonic case ($R_\Omega=0.39$); (b): anti-cyclonic case ($R_\Omega=-0.6$). Dotted line: laminar profile. Continuous line: Busse solution eq.(\protect\ref{Busse}).}
\label{fig:profil-V}
\end{figure}

For the turbulent flow following the subcritical transition, we use the data of Richard~\cite{Rich01}, collected for different Reynolds numbers and rotation numbers. Figure~\ref{fig:profil-V} displays typical turbulent mean profiles in both the cyclonic and anticyclonic cases, for comparison with the laminar and the Busse profiles.

\begin{figure}[h!]
\includegraphics[width=7.5cm,height=5.5cm]{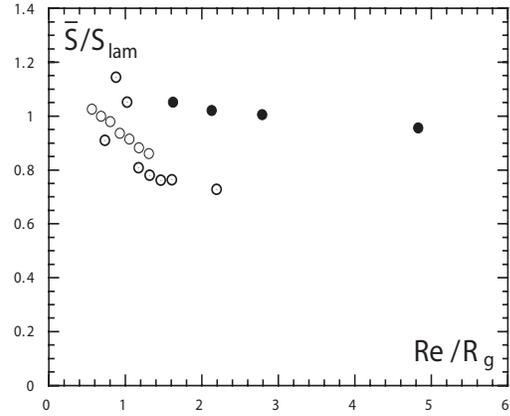}
\caption[]{Ratio of the turbulent mean shear to the laminar shear variation with the Reynolds number (data from Richard).
Cyclonic case: $\odot$: $R_\Omega=0.39$; $\circ$: $R_\Omega\in[0.64, 0.52]$.
Anticyclonic case: $\bullet$: $R_\Omega\in[-2, -1]$.}
\label{fig:shear-redu}
\end{figure}

One notices the profile tendency to evolve from the laminar one to the Busse solution, even if they are still very far away from the extremizing solution. In order to evaluate how fast the convergence occurs, figure~\ref{fig:shear-redu} displays the ratio of the turbulent mean shear to the laminar shear, both estimated at $\tilde r$, i.e. ${\bar S}(\tilde r)/{\tilde S}$, as a function of the ratio of the Reynolds number to the threshold for shear sustained turbulence i.e. $Re/R_g$. One may indeed observe a tendency of shear reduction as the Reynolds number increases, with a more rapid reduction for rotation number closer to $0$. However, none of the case studied by Richard  approaches the value $0.25$ predicted by Busse. It would be interesting to conduct higher Reynolds number experiments at large value of the rotation number, to check whether rotation merely slow down the convergence towards the $0.25$ value, or change it into a number depending on the rotation number.

\begin{figure}[hhh]
\centering
\includegraphics[width=7.5cm]{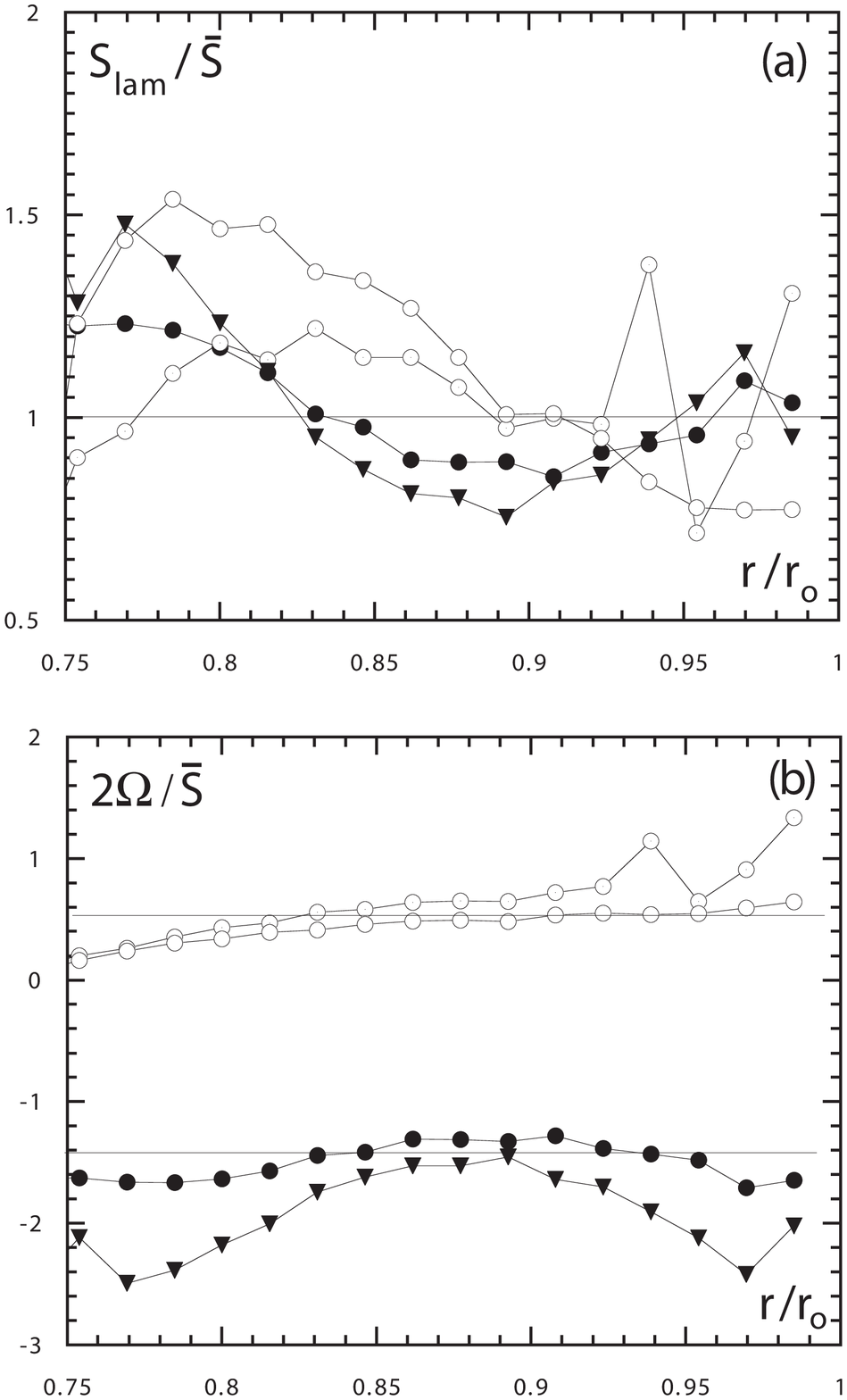}
\caption[]{Mean profile at $Re/R_g=1.6 $ for (a) turbulent transport, traced by the ratio $S^{lam}/S$; (b) local rotation number. $\blacktriangledown$: $R_\Omega=-1.31$; $\bullet$: $R_\Omega=-1.41$; $\circ$: $R_\Omega=0.39$; $\odot$: $R_\Omega=0.51$. Data are from Richard.}
\label{fig:mean-prof}
\end{figure}

Also one may notice that the decrease of ${\bar S}(\tilde r)/{\tilde S}$ with $Re/R_g$ is much faster for cyclonic flows than for anticyclonic ones. Figure~\ref{fig:mean-prof} may provide some hints on the origin of this dissymmetry. The first one is obtained by studying the radial variation of the ratio $S^{lam}/{\bar S}$ at a given $Re/R_g$, for different rotation number. This quantity provides the radial variation of the turbulent viscosity and thus is a good tracer of transport properties. One may observe an interesting tendency for cyclonic flow to display enhanced (resp. depleted) transport at the inner (resp. outer) core boundary, while anti-cyclonic flow rather displays depleted transport at the center, and enhanced transport at both boundaries.

The second one is provided by the function:
\begin{equation}
{\bar q}(r)=\frac{2{\bar \Omega}(r)}{{\bar S}(r)}=\frac{d\ln{\bar\Omega}}{d\ln r},
\label{qlocal}
\end{equation}
which may be viewed either as a local mean angular velocity exponent, or a local mean rotation number. This local exponent also plotted on figure~\ref{fig:mean-prof}, for different rotation number, at $Re/R_g=1.6$. One clearly observes a tendency towards constancy of this local exponent in the core of the flow and a bimodal behavior: cyclonic flow scatters towards $q=0.5$ while anti-cyclonic flow scatters towards $q=-1.5$. We have observed a persistence of this behavior at larger Reynolds number (up to at least $Re/R_g=20$).

\section{Torque measurements and transport properties}

The turbulent transport can be estimated via the torque $T$ applied by the fluid to the rotating cylinders. Traditionally, one works with the non-dimensional torque $G=T/\rho h\nu^2$~\cite{LFS92}. For laminar flows, one can compute this torque analytically using the laminar velocity profile. It varies linearly with the Reynolds number.
\begin{equation}
G^{lam}=\frac{2\pi}{(1-\eta)^2}\eta Re.
\label{torquelam}
\end{equation}
When the turbulence sets in, the torque applied to the cylinders tends to increase with respect to the laminar case. A good indicator of the turbulent transport can then be obtained by measuring $G/G^{lam}$.

\subsection{Super-critical case}
\label{torque-super}

As noticed by Richard and Zahn~\cite{RZ99}, most of the torque measurements available in the literature concern the case with the outer cylinder at rest (see e.g.~\cite{LFS92,LewiSwin99} and references therein). In that case, we note that $\vert R_\Omega\vert =\vert \eta-1\vert \le 1$. An example of the variation of $G/G^{lam}$ with Reynolds number is given in Figure~\ref{fig:torque}, in an apparatus with $\eta=0.724$. One observes three types of behaviors: below a Reynolds number $R_c$, i.e. in the laminar regime $G/G^{lam}=1$. Above $R_c$, one observes a first regime in which $G/G^{lam}$ varies approximately like a power-law, with exponent $1/2$. In this regime, Taylor vortices can often be noticed. This regime continues until $Re=R_T$, where the torque becomes stronger, and the power-law steepens into something with exponent closer to $1$. This regime has been observed up to the highest Reynolds number achieved in the experiment (of the order of $10^6$).

The experiment with inner cylinder rotating only covers flows such that $R_\Omega=\eta-1$. To check whether this kind of measurement is  typical of torque behaviors in the globally supercritical case, one must rely on experiments in which the outer cylinder is also in rotation. Unfortunately, the only torque measurements available in this case are quite older~\cite{Wen33} and not as detailed as in the case with inner cylinder rotating. Most specifically, they do not extend all the way down to the transition region between laminar and turbulent. In several instances in which large Reynolds number are achieved, however, one may observe a steepening of the relative torque towards the $G/G^{lam}\sim Re$ already observed in the case with inner cylinder rotating. On other measurements performed at lower Reynolds numbers, the relative torque displays a behavior more closely related to the intermediate regime, with $G/G^{lam}\sim Re^{1/2}$. Altogether, this is an indication that in the globally supercritical case, the torque follows three regimes:
\EQA
G&\sim& a Re,\quad Re<R_{c},\nonumber\\
G&\sim &\beta^{sup} Re^{3/2},\quad R_{c}<Re<R_{T},\nonumber\\
G&\sim& \gamma^{sup} Re^{2},\quad Re>R_{T},\nonumber\\
\alpha&=&\frac{2\pi\eta}{(1-\eta)^2},
\label{torquelinear}
\ENA
where $\beta^{sup}$ and $\gamma^{sup}$ are constants to be specified later

\subsection{Sub-critical case}
\label{torque-sub}

\begin{figure}[hhh]
\centering
\includegraphics[width=7.5cm]{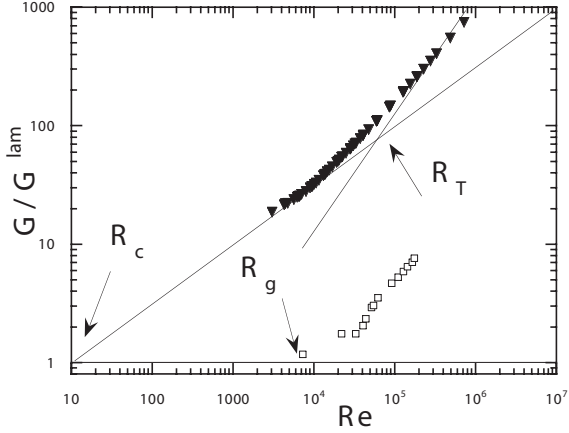}
\caption[]{Relative torque $G/G^{lam}$ as a function of the Reynolds number. $\blacktriangledown$: super-critical case with outer cylinder at rest; $R_\Omega=-0.276$; $\eta=0.724$. Data are from Lewis and Swinney. $\square$: sub-critical case, with inner cylinder at rest. $R_\Omega=0.47$; $\eta=0.68$. Data are from Wendt.}
\label{fig:torque}
\end{figure}

The only measurements of torque in the subcritical case were performed by Wendt~\cite{Wen33} and Taylor~\cite{Tay36} in experiments with the resting inner cylinder, and rotating outer cylinder. Wendt's experiments cover three different values of $\eta$, Taylor's cover eleven values of $\eta$. Taylor measurements cover sufficiently small value of Reynolds number so that one can see that above a critical Reynolds number $Re=R_{g}$, the torque bifurcates from the laminar value towards a regime in which the relative torque $G/G^{lam}$ behaves like $Re$. An example is given in figure~\ref{fig:torque}. Measurements by Wendt at larger Reynolds number display no evidence for an additional bifurcation. So, in the subcritical case, the torque presumably follows only two regimes:
\EQA
G&\sim& a Re,\quad Re<R_{g},\nonumber\\
G&\sim& \gamma^{sub} Re^{2},\quad Re>R_{g},
\label{torquenonlinear}
\ENA
where $\gamma^{sub}$ is a constant that we specify in the next subsection.

\subsection{Connecting torque and thresholds}
\label{torque-thres}

\begin{figure}[h!]
\centering
\includegraphics[width=7.5cm]{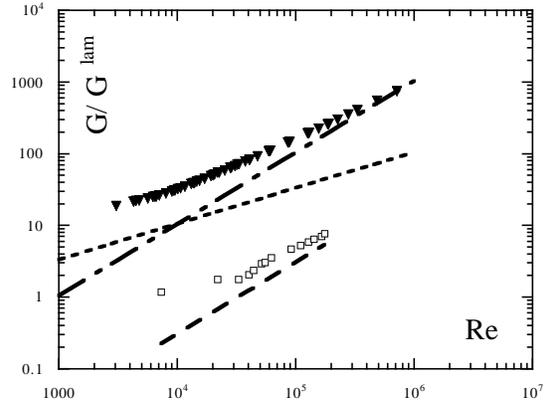}
\caption[]{Relative torque $G/G^{lam}$ as a function of the Reynolds number, compared with its determination using critical Reynolds numbers. $\blacktriangledown$: globally super-critical case with outer cylinder at rest. $R_\Omega=-0.276$; $\eta=0.724$. (Lewis and Swinney). $\square$: globally sub-critical case, with inner cylinder at rest. $R_\Omega=0.47$; $\eta~=~0.68$ (Wendt 1933). Short dashed line: $\sqrt{Re/R_{c}}$ with $R_{c}=90$; dot-dashed line: $Re/\sqrt{R_c R_T}$ with $\sqrt{R_c R_T}=957$; long-dashed line: $Re/R_{g}$ where $R_{g}=32688$. The critical Reynolds numbers have been computed using results of section \ref{stab}.}
\label{fig:test-torque}
\end{figure}

Invoicing the continuity of the torque as a function of the Reynolds number at the transitions allows to determine the prefactors $\beta^{sup}$, $\gamma^{sup}$ and $\gamma^{sub}$. In the supercritical case, one obtains:
\EQA
\beta^{sup}&=&\alpha R_{c}^{-1/2},\nonumber\\
\gamma^{sup}&=&\beta^{sup} R_{T}^{-1/2}=\frac{a}{\sqrt{R_{c}R_{T}}},
\label{relationslinear}
\ENA
and in the subcritical case:
\begin{equation}
\gamma^{sub}=\frac{a}{R_{g}}.
\label{relationnonlin}
\end{equation}
where $\alpha$ is known through (\ref{torquelam}). This enables the knowledge of the torque as a function of $R_c$ and $R_T$, or $R_g$ which then encode all the dependencies on $R_\Omega$ and $\eta$. This would be of great practical interest, and a posteriori gives all its importance to the work conducted in section \ref{stab}, since torque measurements are usually more difficult to perform than thresholds estimations, especially when both cylinders are rotating. Our argument is admittedly very crude, so it is important to test its validity on available data. Figure~\ref{fig:test-torque}  shows the comparison between the real non-dimensional torque measured in experiments, and the torque computed using only the critical Reynolds number. At low Reynolds number, there is a fairly large discrepancy but at large Reynolds, the approximate formula provides a good estimate.\

\begin{figure}[hhh]
\centering
\includegraphics[width=7cm, height=6cm]{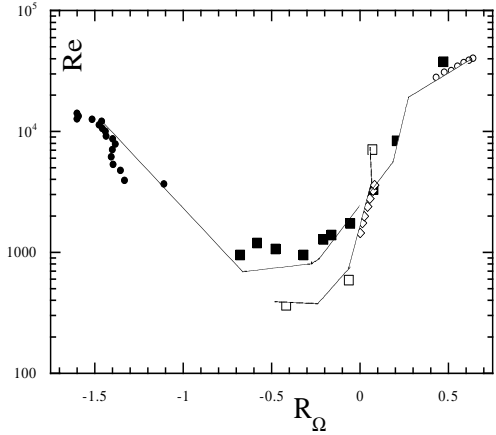}
\caption{$R_{g}$ as a function of $R_\Omega$ and $\eta$. Anticyclonic side:
$\bullet$: Richard data ($\eta=0.7$); $\blacksquare$: Wendt data ($\eta=0.68$). Cyclonic side: $\lozenge$: Tilmark data ($\eta=1$) and fit as in section \ref{stab}; $\square$: Wendt data ($\eta=0.935$); $\circ$: Richard data ($\eta=0.7$). The lines are guides for the eyes to underline the continuity across the supercritical to subcritical domains for similar values of $\eta$.}
\label{fig:Rgcontinuity}
\end{figure}

Comparing (\ref{relationslinear}) and (\ref{relationnonlin}) suggest to introduce ${R_g}^{sup}=\sqrt{R_c R_T}$. This new threshold, defined in the supercritical case, would correspond to the Reynolds number above which turbulence is sustained by the shear mechanism, and not anymore by the linear instability mechanisms. A physical basis for this expression could be given using the observation that the transition occurs in a turbulent state, where transport properties are augmented with respect to a quiescent, laminar case, in which all transport is ensured by viscous processes. This results in a {\sl delayed} transition to the ultimate state, since the viscosity is artificially higher by an amount $\nu_t/\nu$, where $\nu_t$ is the turbulent viscosity. Using $\nu_t/\nu=G/G_{lam}$, we thus get from (\ref{torquelinear}) and (\ref{relationslinear}) an estimate of the relevant threshold as:
\begin{equation}
R_{g}^{sup}=R_T\frac{\nu}{\nu_t}=R_{T}\frac{a}{\beta^{sup}\sqrt{R_{T}}}=\sqrt{R_{c}R_{T}}.
\label{relationreynolds}
\end{equation}
At this stage of the analysis, $R_g^-$, $R_g^{sup}$ and $R_g^+$ respectively define a function of $R_\Omega$ and $\eta$ on the intervals $R_\Omega<{R_\Omega^c}^{-}$, ${R_\Omega^c}^{-}<R_\Omega<{R_\Omega^c}^{+}$ and ${R_\Omega^c}^{+}<R_\Omega$, where we recall here that ${R_\Omega^c}^{-}=-1$ and ${R_\Omega^c}^{+}=1/\eta-1$. Further indication of the relevance of $R_g^{sup}$ is provided by the continuity of this function with $R_\Omega$ throughout the super/sub-critical boundaries. This is illustrated on figure~\ref{fig:Rgcontinuity}, where the continuity is obtained on the cyclonic side between the Tillmark's data ($\eta=1$) and Wendt's data ($\eta=0.935$) and on the anticyclonic side between Richard's data ($\eta=0.7$) and Wendt's data ($\eta=0.68$).

\subsection{How to use torque with resting outer cylinder}
\label{torque-out}

\begin{figure}[hhh]
\centering
\includegraphics[width=7cm, height=6cm]{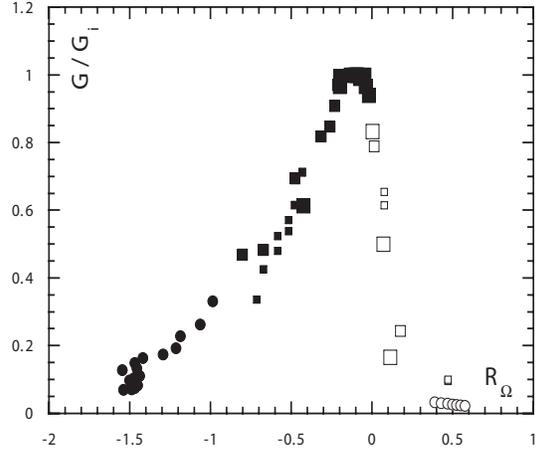}
\caption[]{Relative torque $G/G_i$ as a function of $R_\Omega$ and $\eta$. 
$\bullet$, (resp. $\circ$): estimation from Richard data ($\eta=0.7$), based on critical Reynolds numbers, computed using results of Section \ref{stab} in the anticyclonic (resp. cyclonic case); $\blacksquare$, (resp. $\square$): Wendt data ($\eta=0.68, 0.85, 0.935$), the square size increasing with $\eta$ in the anticyclonic (resp. cyclonic case).}
\label{fig:relative-torque}
\end{figure}

Torque measurements described in previous section suggests that at large enough Reynolds number, an "ultimate" regime is reached with quadratic variation with Reynolds number. This suggests that {\it in this regime}, the interesting parameter is the ratio of the torque in any configuration, to the torque measured in a special case. Because the case with resting outer cylinder is the most studied, it is of practical interest to choose this case as the reference, so that the relevant ratio is $G/G_i$, where $G_i$ is the torque when only the inner cylinder is rotating. Given the above subsections, $G/G_i$ is only a function of $R_\Omega$ and $\eta$ given by $h(R_\Omega,\eta)=R_g(R_\Omega(R_o=0),\eta)/R_g(R_\Omega,\eta)$, where $R_g$ is the generalized threshold defined in the previous section and displayed on figure~\ref{fig:Rgcontinuity}. Figure~\ref{fig:relative-torque} indeed shows the ratio $G/G_i$, for different values of $\eta$, as a function of the rotation number. The measurements for $-0.8<R_\Omega<0.5$ are direct measurements from the Taylor and Wendt experiments. The measurements for $R_\Omega<-1$ and $R_\Omega>0.5$ are indirect measurements, coming from the experiment by Richard, in which only critical numbers from stability were deduced. In that case, the torques have been computed using the results of previous section. All these results show that the non-dimensional torque behaves as: 

\begin{equation}
G(Re, R_\Omega,\eta)=G_i(Re,\eta) h(R_\Omega,\eta),
\label{previous}
\end{equation}
where $G_i$ is the torque when only the inner cylinder is rotating, and $h(R_\Omega,\eta)$ is the function of figure~\ref{fig:relative-torque}.

This universal function is very interesting because it provides good insight about the influence of the rotation and curvature on the torque. For rotation number $-0.2<R_\Omega<0$, the torques are maximal and equal to the torque measured when only the inner cylinder is rotating. For rotation numbers outside this range, torques tend to decrease, with a sharp transition towards a constant of the order of $0.1$ on the side $R_\Omega>0$. On the other side, the transition is softer, with an approximate quadratic inverse variation  until the smallest available rotation number $R_\Omega=-1.5$. From a theoretical point of view, the asymmetry could be linked with the different stability properties of the flow on either side of the curve: for $-1<R_\Omega<1/\eta-1$, the flow is linearly unstable, while it becomes liable to finite amplitude instabilities outside this range. The variation we observe can also be linked with experimental studies by Jacquin et al~\cite{Jacq90}, re-analyzed by Dubrulle and Valdettaro~\cite{DubrVald92}.  They show that rotation tends to inhibit energy dissipation and observed simple power laws linking the energy dissipation with and without rotation as $\epsilon_{\Omega\neq 0}=\epsilon_{\Omega=0} R_\Omega^{\gamma}$, where $R_\Omega$ is a rotation number based on local shear and rotation.\

\begin{figure}[hhh]
\centering
\includegraphics[width=7.5cm]{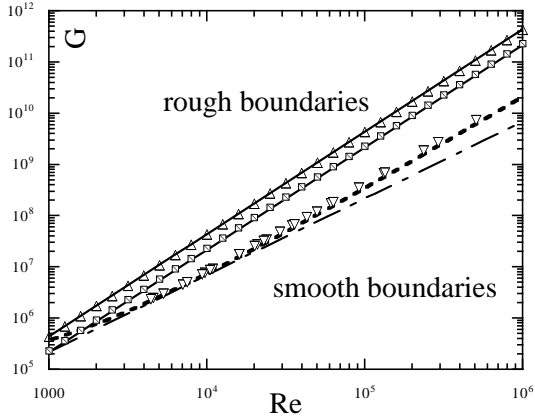}
\caption[]{Influence of boundary conditions on torque. Case with two rough boundaries at $\eta=0.724$, $\triangle$, data from Van den Berg et al.; at $\eta=0.625$, $\boxplus$, data from Cadot et al.. The continuous lines are the formula (\protect\ref{torquerough}). Case with two smooth boundaries at $\eta=0.724$, $\triangledown$. Data from Lewis and Swinney. The dotted and the dashed-dot lines are the formulae (\protect\ref{modelDH}).}
\label{fig:torque-rough.eps}
\end{figure}

Finally, the previous discussion shows that the knowledge of the torque in the case with resting outer cylinder as a function of $Re$ and $\eta$ is an essential data to compute the torque in any other configuration.  A theoretical model of the torque in that configuration has been proposed by Dubrulle and Hersant~\cite{DubrHers02}, in the case where the boundary conditions at the cylinder are smooth. It gives:
\EQA
G_i&=& \frac{(3+\eta)^{1/4} (\eta Re)^{3/2}}{(1-\eta)^{7/4}(1+\eta)^{1/2}},
\quad R_{c}\le Re\le R_{T}
\nonumber\\
G_i&=&0.33\frac{(3+\eta)^{1/2}}{(1-\eta)^{3/2}(1+\eta)}\frac{(\eta Re)^2}{\left(\ln[(K(\eta)(\eta Re)^2]\right)^{3/2}},\nonumber\\
&&\quad\quad\quad Re>R_{T},\nonumber\\
K(\eta)&=& 0.0001\frac{(1-\eta)(3+\eta)}{(1+\eta)^2},
\label{modelDH}
\ENA
The quality of the fit can be checked on Fig. \ref{fig:torque-rough.eps}. For $Re<R_{c}$, the flow is laminar and the transport is ensured only by the ordinary viscosity. 

\subsection{Towards extended Reynolds similarity}
\label{torque-ext}

The link between torque and critical Reynolds number has a powerful potential for generalization of the torque measurements performed in the laboratory for astrophysical or geophysical flows. Indeed, all the additional complications studied so far (aspect ratio, circulation, magnetic field, stratification, wide gap limit) have been found to shift the critical Reynolds number for linear stability by a factor function of this effect, like $R_{c}(effect\neq 0)=R_{c}(effect=0) F$. Depending on the situation, $F$ can be interpreted as either a change in the effective viscosity (magnetic field), or a change in the effective length scale (aspect ratio, wide gap). If, on the other hand, the scaling of the torque with Reynolds number (i.e. the shear) remains non-affected by such a process, the computation done in section \ref{torque-out} are easy to generalized through an {\sl effective Reynolds} number $Re^{eff}=Re/F$. Specifically, everything that has been said for the torque, in the ideal Taylor-Couette experiment, will still be valid with additional complication provided one replaces the Reynolds number by an effective Reynolds number, taking into account the stability modification induced by this effect. This principle is by no mean trivial and must be used with caution, even though it may appear as nothing more that an extension of the Reynolds similarity principle. In fact, it has been validated so far only in the case with vertical magnetic field, where it has been indeed checked by Donnelly and Ozima~\cite{DonnOzim62} that the torque scaling is unchanged by the magnetic field. In the sequel, we shall use this procedure in disks, because we noticed that it gave the most sensible results. It would however be important to check experimentally this "extended Reynolds similarity" principle. 

\subsubsection{Influence of boundary conditions}
Experimental investigation of the Taylor-Couette flow with different set-up has shown that boundary conditions have an influence on the torque. More precisely, it has been shown that the inclusion of one~\cite{BDLL03} or two~\cite{CCDDT97} rough boundary condition, in configuration with outer cylinder at rest, increases the torque with respect to the case with two smooth boundary conditions, at large Reynolds numbers. In convective flows, a similar increase of transport properties is observed when changing from no-slip to stress-free boundary conditions~\cite{Werne96}. In both cases, the increase occurs so as to increase the agreement between the observed value, and a value based on classical Kolmogorov theory. A theoretical study of Dubrulle~\cite{Dubr01} explains this feature through the existence or absence of logarithmic corrections (see formula \ref{modelDH}-b)) to scaling generated by molecular viscosity and large-scale velocity gradient in the vicinity of the boundary. Obviously, in the presence of a rough boundary, or under stress-free boundary conditions, mean large-scale velocity gradients are erased near the boundary, and no logarithmic correction develops.\

For two rough boundary conditions, Cadot et al~\cite{CCDDT97} measure $G_i\sim 0.22-0.3 Re^2$ for $\eta=0.625$, while van den Berg et al.~\cite{BDLL03}observe $G_i\sim 0.43 Re^2$ for $\eta=0.73$. The analogy with thermal convection~\cite{DubrHers02} suggest that $G_i$ depends on its laminar value and on $\eta$ and $Re$ like
\begin{equation}
\frac{G_i}{G_{lam}}=\gamma_{rough} \frac{\sqrt{(3+\eta)(1-\eta)}}{(1+\eta)}\eta Re.
\label{anlconv}
\end{equation}
Using (\ref{torquelam}), and the experimental law, we find $\gamma_{rough}=0.017$, so that:
\begin{equation}
G_i=0.107 \frac{\sqrt{(3+\eta)}}{(1+\eta)(1-\eta)^{3/2}}(\eta Re)^2.
\label{torquerough}
\end{equation}
The comparison between this formula and the experiments is made in figure~\ref{fig:torque-rough.eps}. For reference, we also added the torque in the case of two smooth boundary conditions, as given by (\ref{modelDH}-b).\

We do not have any theory for the case with asymmetric boundary conditions (one rough, one smooth). Laboratory experiments show that the torque lies in between the curve for two smooth boundary conditions and the curve for two rough boundary conditions. The exact location however depends on local conditions in a non-trivial way (for example it is different when the rough conditions applies to the (rotating) inner cylinder or to the (resting) outer cylinder). The present experimental evidence therefore only allows the torque measurements with two smooth (resp. two rough) boundary conditions to be considered as lower (resp. upper) bounds for the torque, in case of complicated boundary conditions.

\subsubsection{Influence of radial circulation}
\label{torque-rad}
\begin{figure}[h!]
\centering
\includegraphics[width=8cm]{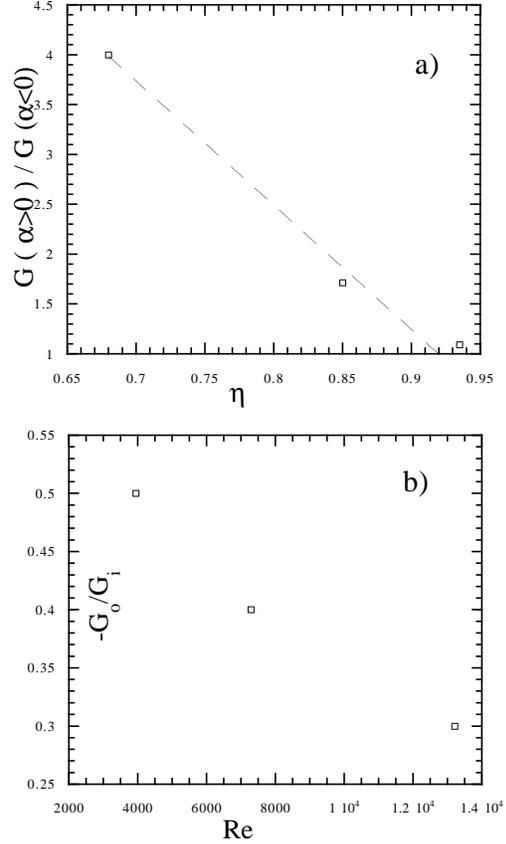}
\caption[]{Influence of radial circulation on torque. (a) $G(\alpha>0)/G(\alpha<0)$ as a function of $\eta$ for $\Omega_0/\Omega_i=0$ in the experiment of Wendt, at large Reynolds numbers. The squares are the data. The line is the fit eq. (\protect\ref{fitWE2}) (b) Ratio of torque applied to outer cylinder vs. torque applied to the inner cylinder in the case of an outward circulation, $\eta=0.89$.}
\label{fig:rad-torque.eps}
\end{figure}

Torque measurements by Went, for different geometry show that the circulation can have an influence on the transport properties. Specifically, it has been observed that an outward circulation tends to increase the torque applied on the inner cylinder, while an inward circulation tends to decrease this torque. The difference can be quite important. At large Reynolds numbers, the relative increase of the torque $G(\alpha<0)/G(\alpha>0)$ can be computed as a function of $\eta$. This is shown in Fig. \ref{fig:rad-torque.eps}. One observes a quasi-linear variation:
\begin{equation}
\frac{G(\alpha<0)}{G(\alpha>0)}=12.45(1-\eta), \quad\Omega_i=0.
\label{fitWE2}
\end{equation}
The case with intermediate boundary conditions (presumably $\alpha$ close to zero) lays about half way in between the two cases so that:
\begin{equation}
\frac{G(\alpha<0)}{G(\alpha=0)}=4.75(1-\eta), \quad\Omega_i=0.
\label{fitWE3}
\end{equation}

Close to the transition threshold, there is also an asymmetry between the two circulation regimes: outward circulation enhances the torque with respect to the laminar regime, while inward circulation decreases this torque! This puzzling aspect has been explained by Coles and Van Atta~\cite{CvA66}; in absence of circulation, in stationary state, the torque exerted at the inner and outer cylinder must balance. In the presence of circulation, the transport of fluid toward or away from the plane of symmetry induces an imbalance of the two torques, which ceases to be equal. Coles and Van Atta measured this imbalance as a function of the Reynolds number for the case with inner cylinder at rest, at $\eta=0.89$, and with boundary conditions favoring an outward circulation. One observes an imbalance of the order of 30 to 50 percent on Fig. (\ref{fig:rad-torque.eps}), with the torque on the inner cylinder being larger. These observations suggest the following model: in the presence of a radial circulation, the inner and outer torques are modified into:
\EQA
G_o(\alpha)&=&G(\alpha=0)\left(1-\alpha\Delta(Re,\eta)\right),\nonumber\\
G_i(\alpha)&=&G(\alpha=0)\left(-1-\alpha\Delta(Re,\eta)\right),
\label{modele}
\ENA
where $\Delta$ is a positive function of $\eta$ and $Re$. So, when reversing the circulation (going from $\alpha>0$ to $\alpha<0$, the torque exerted at the inner cylinder decreases (in absolute value), like in Wendt data. Moreover, these data indicate that at large Reynolds number, the function $\Delta$ becomes independent of $Re$. Note also that according to this model, we should have $G_i(\alpha<0)/G_i(\alpha>0)=-G_o(\alpha>0)/G_i(\alpha>0$. At $Re=10^5$, $\eta=0.8$, the data of Wendt provide a value of $0.25$ for this ratio, in good agreement with the value observed by Coles and Van Atta, see Fig. \ref{fig:rad-torque.eps}. In this model, the total torque is zero (conservation of total angular momentum) only when considering the torque applied by the circulation on the top and the bottom boundary. This means that in the presence of a radial circulation, a non-negligible torque is likely to apply at the vertical boundary. This observation may be relevant to astrophysical disks, and jet-like phenomena.

\subsubsection{Influence of a vertical magnetic field}
\begin{figure}[hhh]
\centering
\includegraphics[width=8.5cm]{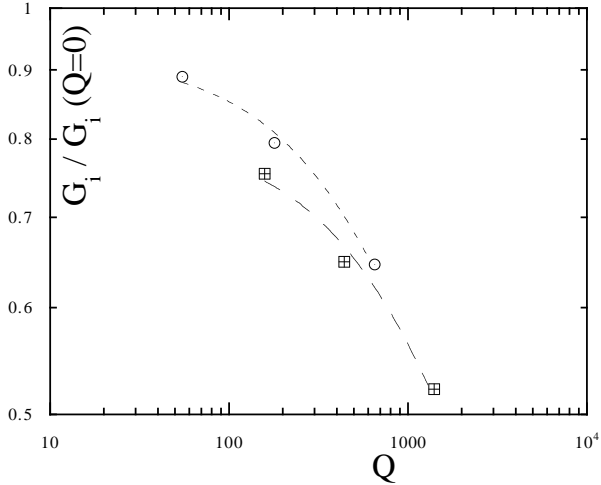}
\caption[]{Influence of vertical magnetic field on torque in the case with inner cylinder rotating.  Ratio of torque with magnetic field to the torque without magnetic field in same experimental conditions, as a function of $Q$, the non-dimensional magnetic number. The symbols are the data. The lines are the fit (\protect\ref{reducmag}). Data are form Donnelly and Ozima. $\boxplus$: $\eta=0.901$, $Re=2105$; the constant for the fit are $b=0.803$ and $c=0.001$; $\odot$: $\eta=0.995$, $Re=2050$; the constants used for the fit are $b=0.920$ and $c=0.002$.}
\label{fig:ozima-torque.eps}
\end{figure}

The influence of a constant vertical magnetic field on the torque has been studied by Donnelly and Ozima~\cite{DonnOzim62}. The measurements have been performed in the linear instability regime, with outer cylinder at rest. It is observed that an increasing magnetic field reduces the torque, so as to conserve the $Re^{3/2}$ scaling observed at zero magnetic field (section \ref{torque-super}). The torque reduction is thus a function only of a non-dimensional magnetic field, and of $\eta$. Examples are provided in Fig.~\ref{fig:ozima-torque.eps}, for gap sizes $0.901$ and $0.995$ and Reynolds number $Re\sim 2100$.

\noindent
The torque reduction can be quantified by the dimensionless number $Q=\mu^2 H^2\sigma d^2/\rho\nu$, where $\mu$ is the permeability, $\sigma$ is the electrical conductivity and $H$ is the magnetic field in tesla. It seems to follow a simple law:
\begin{equation}
\frac{G_i}{G_i(Q=0)}=\frac{b(\eta)}{\sqrt{1+c(\eta) Q}},
\label{reducmag}
\end{equation}
where $b(\eta)$ and $c(\eta)$ are functions of the gap size. Physically, this torque reduction may be due to the elongation of the cellular vortices which occurs as the magnetic field is increased~\cite{DonnOzim62}. Mathematically, the reduction can be understood using the connection between torque and critical number. In this framework,  Chandrasekhar~\cite{Chan60} observes that the addition of a magnetic field onto a flow heated from below imparts to the liquid an effective kinematic viscosity $\nu_{eff}\propto (H_z d)^2/\kappa_m$. Only the component of the field parallel to the gravity vector is effective. This makes the critical Reynolds number for stability proportional to $1+Q/Q_0$. Using the relation between the torque and the critical Reynolds number in the linearly unstable regime (eqs. (\ref{torquelinear}) and (\ref{relationslinear})), this leads to the scaling (\ref{reducmag}).

\subsection{Turbulent viscosity}
\label{torque-visc}

The turbulent viscosity in the direction perpendicular to the shear $\nu_{t}$ can be estimated via the mean torque $\bar T$ applied by the fluid to the rotating cylinders and the mean turbulent velocity profile. Indeed, this torque induces a stress equal to:
\begin{equation}
\frac{\bar T}{rA}=\frac{\rho\nu^2}{2\pi r^2}{\bar G}=\rho \nu_{t}r\partial_r \frac{{\bar u_\theta}}{r},
\label{stress}
\end{equation}
where $A$ is the area of a cylindrical fluid element at radius r, $\rho$ is the fluid density, and ${\bar u_\theta}$ is the mean azimuthal viscosity.  Since a similar formula applies in the laminar case, with $\nu_t=\nu$, one simply gets:
\begin{equation}
\frac{\nu_{t}}{\nu}=\frac{S_{lam}}{{\bar S}}\frac{{\bar G}}{G_{lam}}.
\label{viscositysimple}
\end{equation}
Using the expression of $Re$, $R_{\cal C}$ and $G_{lam}$ (\ref{torquelam}), we thus get the simple expression:
\begin{equation}
\nu_t=\frac{1}{2\pi}{R_{\cal C}}^4\frac{G_i(Re,\eta )}{Re^2}h(R_\Omega,\eta )\frac{S_{lam}}{\bar S}{\tilde S}{\tilde r}^2=\beta{\tilde S}{\tilde r}^2
\label{viscosityradial}
\end{equation}
\noindent
Here, we have adopted the notation of Richard and Zahn~\cite{RZ99} to express the turbulent viscosity in unit of the typical shear and radius of the flow as $\beta$. This non-dimensional parameter encompasses all the interesting variation of the turbulent viscosity as a function of the radial position $r$ and the control parameters $Re$, $R_{\cal C}$ (or $\eta$) and $R_\Omega$.  The radial variation is given through the ratio $S_{lam}/{\bar S}$ as illustrated in section \ref{mean-sub}. This ratio is one near the boundary and may increase in the core of the flow, due to the turbulent shear reduction. All the variation with $R_\Omega$ is through the function $h$ which has been empirically determined in Section \ref{torque-out} and plotted in Fig.~\ref{fig:relative-torque}. All the variation with $Re$ is through $G_i(Re,\eta )/Re^2$, which can be determined through torque measurements (Sections \ref{torque-super} and \ref{torque-sub}), with a theoretical expression provided in Section \ref{torque-out} for smooth boundaries, and \ref{torque-ext} for rough boundaries. The dependence on the curvature is subtler since it appears in all the above dependencies.

\begin{figure}[hhh]
\centering
\includegraphics[width=7.5cm]{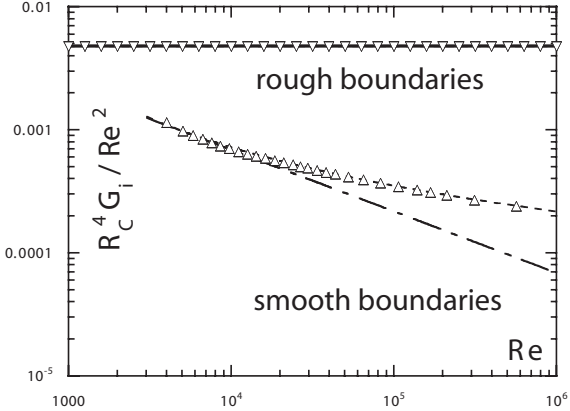}
\caption[]{Influence of Reynolds number on turbulent viscosity. Case with two rough boundaries at $\eta=0.724$, $\nabla$, data from Van den Berg et al.. The continuous line is drawn using formula (\protect\ref{torquerough}). Case with two smooth boundaries at $\eta=0.724$, $\triangle$, data from Lewis and Swinney. The dotted and the dashed-dot lines are drawn using formulae (\protect\ref{modelDH}). }
\label{fig:turb-visc}
\end{figure}

An example of variation of the dimensionless turbulent viscosity $\beta$ for $\eta=0.72$ is provided in Fig.~\ref{fig:turb-visc} for smooth and rough boundary conditions. One sees that at large enough Reynolds number, this function becomes independent of the Reynolds number for rough boundary conditions, while it decreases steadily in the smooth boundary cases, due to logarithmic corrections. This weak Reynolds number variation is in contrast with standard turbulent viscosity prescription, based on dimensional consideration {\sl \`a la Kolmogorov}.\

Finally, let us compare our results with previous results for the turbulent viscosity in rotating flows. Using a turbulent closure model of turbulence, Dubrulle~\cite{Dubr92} derived $\nu_t=2\times 10^{-3} R_\Omega^{-2}{\tilde S}{\bar r}^2$. This formula reflects the correct behavior in term of $R_\Omega$ (see Section \ref{torque-thres}) but fails to reproduce the Reynolds dependence in the case of smooth boundary conditions. For rough boundary conditions, our formula predicts a turbulent viscosity going like $\nu_t=8\times 10^{-3} R_\Omega^{-2}{\tilde S}{\tilde r}^2$ for $R_\Omega<-0.5$ and $\nu_t=2\times 10^{-3} R_\Omega^{-2}{\tilde S}{\tilde r}^2$ for $R_\Omega>0.5\times 10^{-4}$, in the wide gap limit. The formula of Dubrulle is therefore in between these two predictions. 
Richard and Zahn~\cite{RZ99} used Taylor measurements~\cite{Tay36} to derive the value $\beta=1.5\pm 0.5\times 10^{-5}$. These measurements are performed for $Re\sim 5\times 10^4$, with inner cylinder at rest. At $\eta=0.7$, one has $R_\Omega=0.4$ and from Fig.~\ref{fig:relative-torque}, $h(0.4, 0.7)=0.05$. For $S_{lam}/{\bar S}$ we adopt a value equal to $2$, as suggested by Section \ref{mean-sub}. Finally, from Fig. \ref{fig:turb-visc}, we get $R_{\cal C}^4 G_i(Re,0.7)/Re^2=5\times 10^{-4}$ so that we finally obtain $\beta=8\times 10^{-6}$, an estimate close to the one proposed by Richard and Zahn.

\section{Conclusion and Hints for future work}
\label{conc}

The present work provides us with a prescription for the turbulent viscosity hence the turbulent transport for the Taylor-Couette flow. This prediction clearly indicates the dependencies on the Reynolds number and the rotation number. The curvature effect is much trickier to isolate at least with the available data, since it appears in all the terms of the prescription. Especially, on the cyclonic side, where the Rayleigh criterion depends on the curvature, it is impossible without any phenomenological arguments to isolate the curvature effect from the rotation one within the set of data used here. Since we wanted to remain as close as possible to the existing data, we decided not to reproduce any phenomenological arguments in the present paper (for such an analysis see e.g. Longaretti and Dauchot~\cite{PylOD04}).

The introduction of new control parameters, which rely on the dynamical properties of the flow, rather than on its geometry allows us to envision some application of our result to rotating shear flows in general even if one should remain cautious with the details of the boundary conditions. These new control parameters have a rather general ground, but they remain global quantities. It would be interesting to further develop this approach, by introducing local dynamical control parameters, so that in spatially developing flows, one could conduct a local study of the stability properties.

In order to validate the above prescription, it would definitely be necessary to confront it to more experimental data. On the anticyclonic side, in the subcritical regime, only one value of the curvature has been investigated, so that we have very little idea of its influence. In the supercritical regime, an important hypothesis made here was to introduce $R_g^{sup}$ and to relate it to $R_T$ and $R_c$. Also, we have proposed to relate the torque measurements (a difficult experimental task) to the threshold determination. These conjectures should be checked against more data. Finally, we have tried to provide some indications on the influence of external effects such as stratification, or magnetic fields. Clearly the lack of experimental data here is such that very little could be done and a definitive effort should be conducted in this direction.

Still, it is to our knowledge the first time that using most of the existing experimental studies a practical prescription for the turbulent viscosity is proposed. It can certainly be improved, but we believe that, even at the present level, it can already bring much insight into the understanding of some astrophysical and geophysical flows.

\begin{acknowledgements}
We would like to thank L. Mari\'e, F. Hersant and F. Busse for fruitful discussions. 
\end{acknowledgements}



\clearpage
\newpage

\begin{longtable}[c]{ll}

  \caption{Notations\label{notation}}\\
  \endfirsthead

  \hline
  \multicolumn{2}{c}{\textit{Continued on next page}} \endfoot
  \endlastfoot

  \hline
  \multicolumn{2}{c}{\textbf{Superscript and subscript conventions}} \\
  & \\

  $X$ & any flow variable (e.g., component of velocity) \\
  $X^{lam}$ & laminar part of $X$ \\
  $\bar{X}$ & mean part of $X$ \\
  $X'$ & fluctuating part of $X$ \\
  $\tilde{X}$ & typical value of X \\
  $X^\infty$ & relates to inviscid flows \\
  $X^{pCf}$ & relates to Plane Couette flows \\
  $X^{TCf}$ & relates to Taylor-Couette flows \\
  $X^{sub}$ & relates to subcritical flows \\
  $X^{sup}$ & relates to supercritical flows \\
  $X^+$ & relates to cyclonic flows \\
  $X^-$ & relates to anti-cyclonic flows \\
  ${\bm X}= (X_x,X_y,X_z)$ & inertial frame cartesian components of ${\bm X}$\\
  ${\bm X}= (X_r,X_\theta,X_z)$ & inertial frame cylindrical$^{\mathrm{a}}$ components of ${\bm X}$ \\
  ${\bm X}= (X_r,X_\phi,X_z)$ & rotating frame cylindrical$^{\mathrm{a}}$ components of ${\bm
  X}$ \\

  \hline
  \multicolumn{2}{c}{\textbf{Hydrodynamical quantities}} \\
  & \\

  ${\bm x}$ & position vector \\
  ${\bm u}$ & inertial frame velocity vector \\
  ${\bm w}$ & rotating frame velocity vector \\
  $p,\pi$ & fluid pressure, generalized pressure\\
  $\Omega=u_\theta/r$ & angular velocity \\
  $\Omega_{\mathrm{rf}}$ & angular velocity of the rotating frame\\
  $S$ & velocity shear \\
  $S^{lam}; \bar{S}$ & $du_x^{lam}/dy; d\bar{u_x}/dy$ (plane Couette flow)\\
  $S^{lam}; \bar{S}$ & $r d\Omega^{lam}/dr; r d\bar{\Omega}/dr$ (Taylor-Couette flow) \\
  $L=({\bm x}\times {\bm u})_z=r^2\Omega$ & specific angular momentum \\
  $T$ & torque\\
  $G=T/(\rho h \nu^2)$ & adimensionalized torque\\

  \hline
  \multicolumn{2}{c}{\textbf{Geometric and physical quantities}} \\
  \multicolumn{2}{c}{\textit{General}} \\
  & \\

  $\nu$ & kinematic viscosity \\
  $\nu_t$ & turbulent viscosity \\
  $\rho$ & mass density \\

  & \\
  \multicolumn{2}{c}{\textit{Couette flow}} \\
  & \\

  $\pm V$ & bounding plate velocity \\
  $d$ & gap between bounding plates \\

  & \\
  \multicolumn{2}{c}{\textit{Taylor-Couette flow}} \\
  & \\

  $r_{i,o}$ & Inner, outer cylinder radii \\
  $\Omega_{i,o}$ & Inner, outer cylinder angular velocity \\
  $d=r_o-r_i$ & gap \\
  $\eta=r_i/r_o$ & radius ratio (dimensionless measure of the gap) \\
  $h$ & cylinders height\\

  \hline
  \multicolumn{2}{c}{\textbf{Dimensionless quantities}} \\
  \multicolumn{2}{c}{\textit{Standard adimensionalization}} \\
  & \\

  $[ L ] = d$ & unit of length \\
  $[ T ] = d^2/\nu$ & unit of time \\
  $R_{\pm}= \pm V d/\nu$ & Reynolds number of moving plates
  (plane Couette flow) \\
  $Re=R_+-R_-$ & Reynolds number (plane Couette flow) \\
  $R_{i,o}=r_{i,o}\Omega_{i,o} /\nu$ & Reynolds number of rotating
  cylinders (Taylor-Couette flow) \\
  $Re= |R_o-R_i|$ & Reynolds number (Taylor-Couette flow) \\

  & \\
  \multicolumn{2}{c}{\textit{Dynamical adimensionalization}} \\
  \multicolumn{2}{c}{\textit{(Taylor-Couette flow)}} \\
  & \\

  $[ L ] = d$ & unit of length \\
  $[ T ] = \tilde{S}^{-1}$ & unit of time \\
  $Re=\tilde{S}d^2/\nu$ & Reynolds number \\
  $R_\Omega = 2\Omega_{rf}/\tilde{S}$ & rotation number \\
  $R_\Omega^\infty$ & rotation number at marginal stability
  ($Re=\infty$) \\
  $R_{\mathcal C} = d/\tilde{r}$ & curvature number \\

  & \\
  \multicolumn{2}{c}{\textit{Local dynamical ratios}} \\
  \multicolumn{2}{c}{\textit{(Taylor-Couette flow)}} \\
  & \\

  $\Gamma_\nu=\frac{\mathrm{advection-shear}}{\mathrm{viscous\ dissipation}}$
  & Local ``Reynolds'' ratio \\
  $\Gamma_\Omega=\frac{\mathrm{Coriolis}}{\mathrm{advection-shear}}$
  & Local ``rotation'' ratio \\
  $\Gamma_{\mathcal C}=\frac{\mathrm{curvature}}{\mathrm{advection-shear}}$
  & Local ``curvature'' ratio \\

  & \\
  \multicolumn{2}{c}{\textit{Transition Reynolds numbers}} \\
  & \\

  $R_c$ & first supercritical linear transition \\
  $R_g$ & minimal Reynolds number for self-sustained turbulence \\
  $R_T$ & Transition to ``hard'' turbulence (as traced by torques)
  \\

  \hline
  \multicolumn{2}{c}{\textbf{Symbol conventions in graphs}} \\
  & \\

  $\square,\lozenge,\circ,\vartriangle$ & cyclonic data \\
  $\blacksquare,\blacklozenge,\bullet,\blacktriangle$ &
  anti-cyclonic data \\
  $\square,\blacksquare$ & Wendt (1933) data \\
  $\vartriangle,\blacktriangle$ & Taylor (1936) data \\
  $\lozenge,\blacklozenge$ & Tillmark and Alfredsson (1996) data \\
  $\circ,\bullet$ & Richard (2001) data \\
  $\triangledown, \blacktriangledown$ & Lewis and Swinney (1999) data \\

  \hline
  \multicolumn{2}{l}{\footnotesize{${}^{\mathrm{a}}$ The correspondance between cartesian
  and cylindrical axes is ($x \leftrightarrow -\theta, -\phi$), and
  ($y \leftrightarrow r$).}} \\

\end{longtable}
\end{document}